\documentclass[preprints,review,accept,oneauthor,pdftex]{mdpi} 

\firstpage{1} 
\pubvolume{8}
\issuenum{2}
\articlenumber{122}
\pubyear{2022}
\copyrightyear{2022}
\externaleditor{Academic Editor: {Mikhail E. Sachkov}} 
\datereceived{11 January 2022} 
\dateaccepted{10 February 2022 } 
\datepublished{13 February 2022} 
\hreflink{https://doi.org/10.3390/
universe8020122} 

\Title{RR Lyrae and Type II Cepheid Variables in Globular Clusters: Optical and Infrared Properties}

\TitleCitation{{\it Universe} special issue on ``Recent Advances in Pulsating Stars'', M. Marconi and V. Ripepi, eds.}

\Author{{Anupam Bhardwaj} 
~$^{1,2,\dagger,\ddagger, \$}$}

\AuthorCitation{Bhardwaj, A.;}

\address{%
$^{1}$ \quad Center for Large Telescopes, Korea Astronomy and Space Science Institute, Daedeokdae-ro 776, Yuseong-gu, Daejeon 34055, Korea; {anupam.bhardwajj@gmail.com; abhardwaj@kasi.re.kr} 
\\
$^{2}$ \quad INAF-Osservatorio Astronomico di Capodimonte, Via Moiariello 16, 80131 Napoli, Italy}

\firstnote{{EACOA Fellow.}} 
\secondnote{IAU Gruber Foundation Fellow.}
\thirdnote{Marie Skłodowska-Curie Fellow.} 

\abstract{
Globular clusters are both primary fossils of galactic evolution and formation and are ideal laboratories for constraining the evolution of low-mass and metal-poor stars. RR Lyrae and type II Cepheid variables are low-mass, radially pulsating stars that trace old-age stellar populations. These stellar standard candles in globular clusters are crucial for measuring their precise distances and, in turn, absolute ages, and for the calibration of the extragalactic distance scale. Herein, the evolutionary stages of RR Lyrae and type II Cepheids are discussed, and their pulsation properties, including the light curves, color--magnitude and period--amplitude diagrams, and period--luminosity relations in globular clusters at optical and infrared wavelengths are presented. The RR Lyrae visual magnitude--metallicity relation and the multiband period--luminosity--metallicity relations in globular clusters covering a wide metallicity range are also discussed in detail for their application to the RR Lyrae-based distance~scale.}


\begin{document}

\section{Introduction}

Galactic globular clusters (GCs) have traditionally been used as isolated laboratories for our understanding of many aspects of stellar structure, evolution, and dynamics, and for constraining models of stellar populations. These remarkable astronomical objects are compact, gravitationally bound systems of up to a million stars and host some of the oldest stars. As the oldest known stellar systems, the Milky Way GCs provide important clues to its early formation and continuing evolution. In general, GCs are associated with galaxies of all types and masses and their properties exhibit a strong correlation with that of their host galaxies \citep{harris1991, brodie2006}. With ubiquitously old ages comparable to the age of the Universe, the cluster RR Lyrae (RRL) variables also trace the evolution history of their host galaxies and can provide important clues to their formation and evolution.

Variable stars in GCs play an important role in determining their distances and, thus, constraining their intrinsic properties such as the absolute ages, which also place a limit on the minimum age of the Universe \citep{marin2009}. The earliest studies of RRL variables were primarily based on GCs. Following the discovery of variable stars in GCs \citep{pickering1889}, Solon Bailey discovered hundreds of ``cluster variables'' from the Harvard College Observatory, and later separated those as subtypes of RRL variables \citep{bailey1902}. Among the prototypes of different subclasses of population II Cepheids, W Virginis was discovered first \citep{schonfeld1866}, followed by RV Tauri \citep{ceraski1905} and BL Herculis \citep{hoffmeister1929}. The radial velocity determinations of Cepheids by \citet{joy1937} hinted that W Virginis was quite different from the classical or type I Cepheids of similar periods~\citep{wallerstein1984}. Later, \citet{baade1944} showed in his pioneering work that the stellar populations of the galaxies fall into two distinct groups of solar-neighbourhood-like population I and GC-like population II stars. BL Herculis, W Virginis, and RV Tauri fall into the latter group, and are now representative of different subtypes{\footnote{{We} exclusively use ``BL Her'', `` W Vir'', and ``RV Tau'' to refer to the subclasses of type II Cepheids based on their prototype star---BL Herculis, W Virginis, and RV Tauri, respectively.}} of type II Cepheid (T2C) variables. 

The modern time-domain surveys have provided a plethora of data for RRL and T2C variables at optical wavelengths. At the same time, infrared observations have also increased extensively over the past two decades. High-precision photometric and spectroscopic data have also revealed that GCs are not simple stellar populations, and exhibit star-to-star abundance variations in specific elements (see reviews~\citep{gratton2012, bastian2017}). However, the horizontal branch (HB) morphology in GCs is mainly determined by metallicity, and the differences in HB morphologies of similar metallicity GCs have been attributed to the {\it {second parameter problem} 
} \citep{sandage1960}. While several possible solutions have been proposed for the second parameter (see \citep{catelan2009} for details), internal helium variation is one of the main (second) parameters governing the HB morphology \citep{milone2018}. Despite the recent progress in observational and theoretical frameworks, the impact of chemical composition on pulsation properties and the period--luminosity relations (PLRs) of HB RRL and evolved T2Cs is not well-understood at all wavelengths. 

This review summarizes recent progress in studies of RRL and T2Cs in GCs based on multiband photometric data. It is beyond the scope of this article to provide a comprehensive overview of these variables in GCs, and therefore, it will focus on a comparative study of optical and infrared pulsation properties of population II variables. We refer readers to several excellent reviews on RRL and T2C variables available in the literature: \citep{smith1995, wallerstein2002, catelan2004, sandage2006, catelan2009, mcwilliam2011, beaton2018, matsunaga2018, bhardwaj2020}.
This article is organized as follows: the evolutionary and pulsational scenario for RRL and T2Cs is discussed in Section \ref{sec:sec2}. The optical and near-infrared (NIR) pulsation properties of RRL and T2Cs in GCs are discussed in Sections \ref{sec:sec5} and \ref{sec:sec6}, and a brief summary is provided in Section \ref{sec:sec7}.
\section{Evolution of RR Lyrae and Type II Cepheids}
\label{sec:sec2}

\begin{figure}[H]

\includegraphics[width=9. cm]{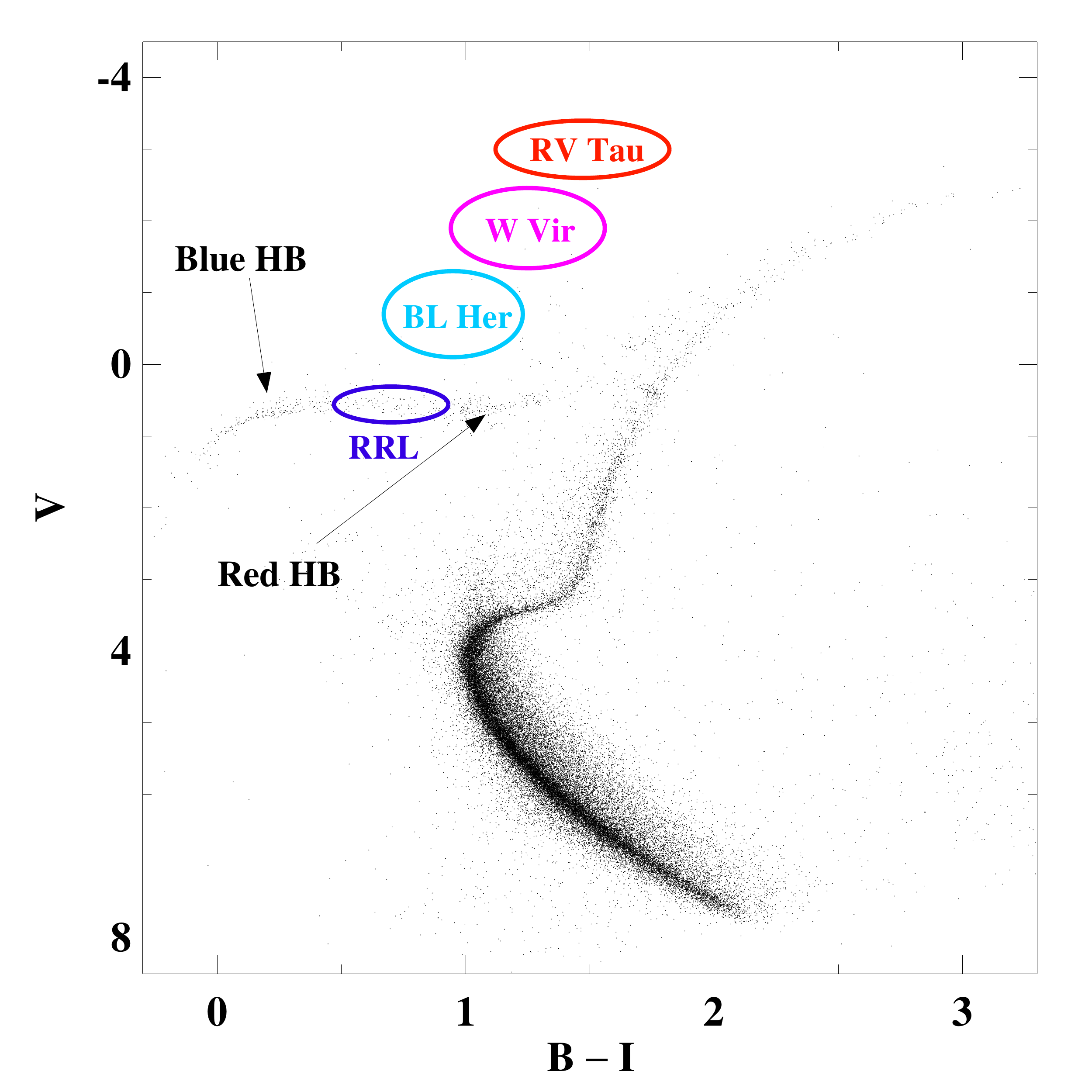}
\caption{{Color}--magnitude diagram for a GC with the approximate positions of RRL, BL Her, W Vir, and RV Tau indicated. The blue and red HB on either side of the RRL ``gap'' are also marked. \label{new_cmd}}

\end{figure}

RRL and T2Cs populate the well-known classical Cepheid instability strip (IS)---a narrow, nearly vertical strip in temperature in the Hertzsprung--Russell (HR) diagram. RRL variables are old ($\geq10$~Gyr), low-mass ($\sim$0.5--0.8$~M_\odot$), metal-poor stars that are located at the cross-section of the HB and the IS. RRL stars pulsate primarily in the fundamental mode (RRab) and first-overtone mode (RRc), with periods in the range between about 0.2~d and 0.8~d. A few RRL variables are classified as doule mode pulsators (RRd) because they pulsate simultaneously in the fundamental and overtone modes (for example, \mbox{\citep{sandage1981, jurcsik2015, soszynski2017}}). T2Cs also belong to old, low-mass, metal-poor stellar populations. A preliminary classification of T2Cs is performed based on their pulsation periods: BL Her ($1\lesssim\!P<5$~d), W Vir ($5\lesssim\!P\!\lesssim20$~d), and RV Tau ($P>20$~d). \citet{soszynski2008a} suggested another subtype, peculiar W Vir (pW Vir, $5\lesssim\!P\!\lesssim10$~d), which are mostly brighter and bluer than W Vir stars. T2Cs primarily pulsate in the fundamental mode, but BL Her, pulsating in the first-overtone mode, have also been discovered \citep{soszynski2019}. {{The} location of RRL and T2Cs on a typical color--magnitude diagram of a GC is shown in Figure~\ref{new_cmd}.}

RRL are in the core helium-burning evolutionary phase of low-mass stars and occupy a region in the HR diagram which is at the intersection between the Cepheid IS and the HB. A low-mass ($\sim$1~M$_{\odot}$) star spends most of its lifetime in the core hydrogen-burning main-sequence phase. After the exhaustion of hydrogen in the core, the inert helium core contracts and the hydrogen burning occurs in a thin shell surrounding the core. The heating of the contracting helium core drives the fusion in the hydrogen shell faster; the star expands as it becomes brighter and redder, moving to the giant branch. At the red giant phase, the inert helium core continues contracting and heating, and the fusion of helium starts at the tip of the red giant in a flash. Once the helium burning in the core becomes the primary source of energy, the star leaves the giant branch as it becomes fainter and bluer. The stellar envelope contracts, attaining the thermal and hydrostatic equilibrium as it reaches the HB phase. The evolution of low-mass stars in the central helium-burning evolutionary phase is discussed in detail in the books of \citet{kippenhahn1991}, \citet{salaris2005}, and \citet{catelan2015}. The zero-age HB (ZAHB) star is characterized by the helium-burning core and the hydrogen-burning shell surrounding the core. The location of ZAHB stars on the HB depends on the total mass (or the envelope mass) for a given helium core mass and envelope composition. For the complex morphology of the HB, the reader is referred to the review by \citet[][and references therein]{catelan2009}. The HB stars have a wide range of effective temperatures, but only stars with specific initial masses ($\lesssim$0.8~$M_\odot$) attain temperatures that place them within the IS. The blue and the red edge of the IS of RRL varies between $\sim$7100~K and $\sim$5200~K, respectively, for a wider range of metal abundances \citep{marconi2015}. The RRL IS shifts to cooler temperatures as the metal abundance increases and becomes systematically fainter for the metal-rich pulsators \citep{bono1997a, marconi2015}.

T2Cs are in the post-HB evolutionary phase of low-mass stars. After the exhaustion of core helium, the inert carbon-oxygen core contracts and heats up in low-mass stars. The helium and hydrogen burning occurs in thin shells outside the core. The star expands rapidly and becomes cooler and brighter reaching the asymptotic giant branch (AGB). It suffers shell flashes at the boundary between the core and the helium region. During these shell flashes, the radius decreases temporarily, and thus, the star moves to the hotter temperatures and inside the IS \citep{gingold1976, wallerstein2002}. T2Cs evolve from the blue HB and reach the IS at higher luminosities than those of RRL. The short-period T2Cs, BL Her, evolve from the HB, bluer than the RRL gap, towards AGB, while depleting core helium. During this process, they cross the IS at luminosities corresponding to a pulsation period of 1 to 5 days. The intermediate-period W Vir evolve to higher luminosities after reaching the AGB and begin to suffer helium shell flashes. This leads to their temporary excursions within the IS at luminosities corresponding to stars with periods between 12 and about 20 days \citep{wallerstein1984}. Once the star becomes more luminous, mass loss reduces the size of the hydrogen shell and it passes through IS once again with periods of more than 20 days, representing the post-AGB evolution of RV Tau variables \citep{bono1997c, wallerstein2002}. However, \citet{groenewegen2017a} suggested that the binarity might be at the origin of the W Vir variables. Similarly, RV Tau stars may also evolve from the more massive and younger objects, or represent binary evolution (see~\citep{groenewegen2017,manick2018} for details).

\section{RR Lyrae Variables}
\label{sec:sec5}

Optical photometric surveys of GCs have been carried out over the past century to investigate their variable star population, including RRL and T2Cs \citep{clement2001}. 
While optical time-series data is plentiful, homogeneous NIR time-series for a statistically significant sample of RRL in a given GC is limited to few studies, which utilized the increasing infrared observations in the last decade. For a comparative investigation of the optical and NIR properties of RRL, the time-series data available in the literature was complied for GCs, covering a wide range of mean-metallicities. Table~\ref{tab1} summarizes the data along with their~references.

\begin{table}[H]

\caption{Source of multiband time-series data for RRL in GCs.}\label{tab1}
\setlength{\tabcolsep}{3.95mm} \resizebox{\linewidth}{!}{\begin{tabular}{ccccccccc}
\toprule
\textbf{GC}& \textbf{[Fe/H]} & \emph{\textbf{D}} & $\boldsymbol{E_{BV}}$ & $\boldsymbol{E_{JK_s}}$ &\textbf{Optical} & $\boldsymbol{N_{\rm{RRL}}}$ &\textbf{NIR} & $\boldsymbol{N_{\rm{RRL}}}$\\
& & \textbf{kpc} & \textbf{mag} & \textbf{mag} & & & & \\
\midrule
M15 & $-2.33$ & 10.71 & 0.10 & --- & $BVI$ \citep{corwin2008} & 89 & $JK_s$ \citep{bhardwaj2021a} & 129 \\
M53 & $-2.06$ & 18.50 & 0.02 & --- & $VI$ \citep{ferro2012} & 64 & $JHK_s$ \citep{bhardwaj2021} & 63 \\
M22 & $-1.70$ & 3.30 & --- & 0.18 & $VI$ \citep{soszynski2019a} & 28 & $K_s$ \citep{alonsogarcia2021} & 62 \\ 
$\omega$ Cen& $-1.64$ & 5.43 & 0.11 & --- & $BVI$ \citep{braga2016} & 179 & $JHK_s$ \citep{braga2018} & 182 \\ 
M3 & $-1.50$ & 10.18 & 0.01 & --- & $BVI$ \citep{jurcsik2015, jurcsik2018} & 148 & $JHK_s$ \citep{bhardwaj2020a} & 228 \\ 
M5 & $-1.33$ & 7.48 & 0.03 & --- & $VI$ \citep{ferro2016} & 28 & $JK_s$ \citep{coppola2011} & 76 \\ 
M4 & $-1.18$ & 1.85 & 0.35 & --- & $BVI$ \citep{stetson2014} & 44 & $JHK_s$ \citep{stetson2014} & 44 \\ 
N6569 & $-0.75$ & 10.53 & --- & 0.25 & $VI$ \citep{soszynski2019a} & 21 & $K_s$ \citep{alonsogarcia2021} & 11 \\ 
N6441 & $-0.45$ & 12.73 & --- & 0.18 & $VI$ \citep{soszynski2019a} & 37 & $K_s$ \citep{alonsogarcia2021} & 23 \\ 
\bottomrule
\end{tabular}}

\noindent{\footnotesize{{\it Notes:} The mean-metallicities ([Fe/H]) and the distance ($D$) to GCs are taken {from} 
\citet{carretta2009} and \citet{baumgardt2021}, respectively. The color-excess values ($E_{BV}=E(B-V)$ and $E_{JK_s}=E(J-K_s)$) are adopted from the references provided with optical and NIR data.}}

\end{table}

\subsection{Multiband Light Curves}

\begin{figure}
\includegraphics[width=13.5 cm]{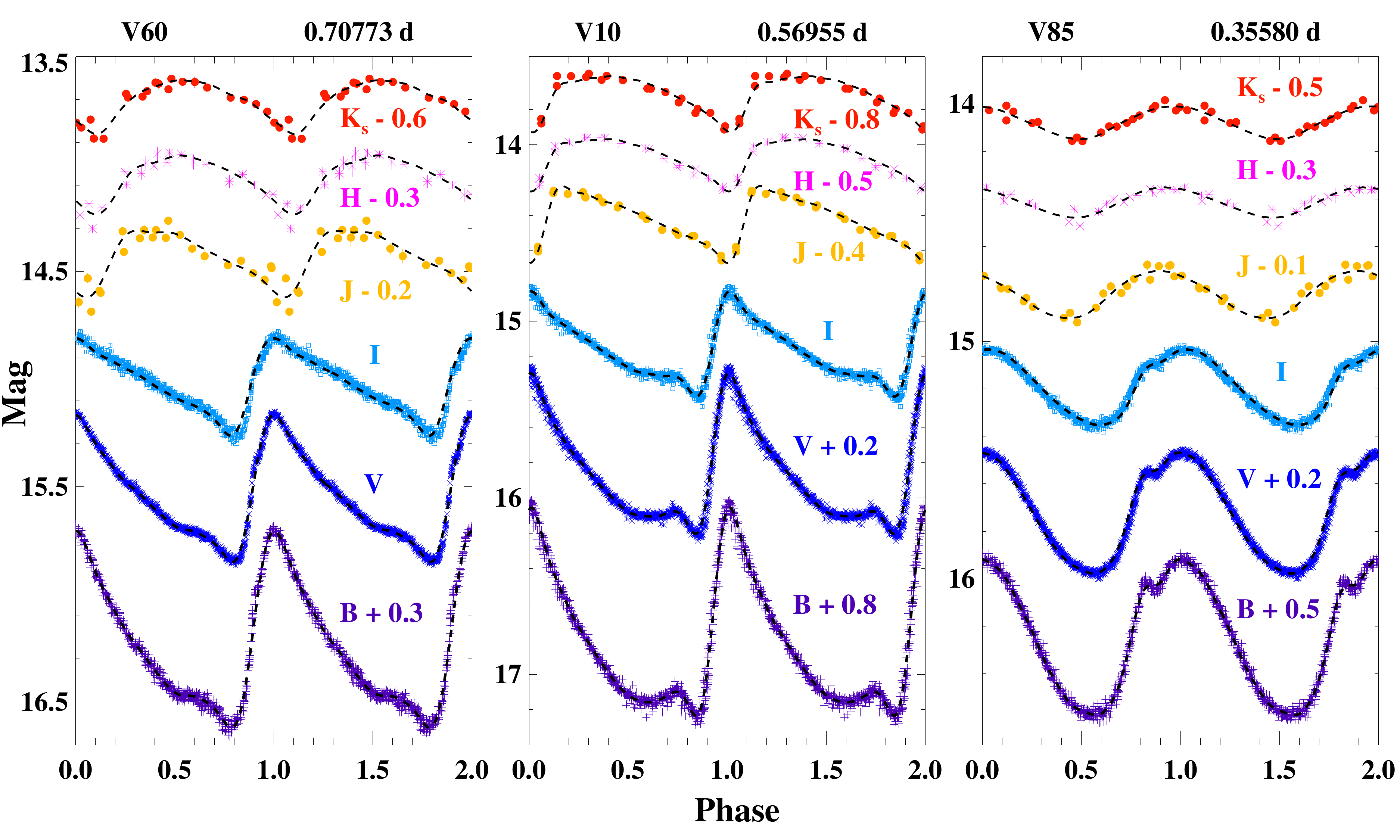}
\caption{Multiband light curves of RRab (left and middle) and RRc (right) stars in M3. A monotonic drop in amplitude is seen as a function of increasing wavelength. The shape of RRab light curves also changes from a saw-tooth shape at optical wavelengths to more sinusoidal shape in infrared bands. In case of RRab, a clear variation in the phase of maximum light is also noted, which shifts toward later phases in infrared bands. The variable star ID and the pulsation periods from the catalog {of} 
\citet{clement2001} are listed at the top of each panel. \label{rrl_lcs}}
\end{figure}

The light curve structure of variable stars is not only crucial for their identification and classification but also for gaining valuable information about the underlying physical processes producing the changes in their brightness. Figure~\ref{rrl_lcs} shows the multiband light curves of RRL in M3. A clear decrease in the amplitudes of the light curves can be seen moving from optical to NIR wavelengths. The skewness and acuteness of RRL light curves attain a value close to unity at infrared wavelengths, implying symmetric sinusoidal brightness variations. The shape of RRab light curves is asymmetric, with a steep rise from minima to maxima and a slow decrease of brightness from maxima to minima at optical wavelengths. Most RRab light curves have a saw-toothed shape with a bump in the vicinity of the minimum brightness and are easily identified. For RRab stars, the phase of maximum brightness shifts significantly to later phases at longer wavelengths. This phase lag in the maximum brightness between optical and infrared bands is similar to fundamental-mode classical Cepheids \citep{madore1991, macri2015, bhardwaj2015}, whereas the phase lag is close to zero for overtone-mode variables. The overtone-mode RRc stars have more symmetric light curves and often exhibit near-sinusoidal variations, even at optical wavelengths. In some cases, RRc stars also exhibit a small bump along the ascending branch. The smaller amplitudes, fainter brightness, and more sinusoidal light curves make the identification and classifications of RRc stars relatively more difficult than for RRab stars.

The stellar pulsation models can successfully predict multiband light and radial velocity variations, and the light curve morphology depends on the chemical composition~\citep{bono2000, marconi2015}. The quantification of the light curve structure based on Fourier decomposition \citep{slee1981} or principal component analysis \citep{deb2009} has been used for a rigorous comparison between observations and theory and to provide constraints for the stellar pulsation models \citep{simon1982, bono1996, marconi2005, das2018}. The light curve parameters of RRL stars have been used to obtain photometric metallicity measurements based on an empirical relation correlating period, Fourier phase parameter ($\phi_{31}$), and [Fe/H]~\citep{jk1996, nemec2013}. The application of machine-learning methods on theoretical and observational data have also shown that the light curve parameters play an important role in predicting the physical parameters of RRL variables \citep{bellinger2020}. 
The light curves shown in Figure~\ref{rrl_lcs} display stable and periodic variations, but a significant fraction of RRL also exhibit the Blazhko effect \citep{blazhko1907}. The Blazhko eﬀect results in (multi-)periodic modulations of the light curves on a timescale that is much longer than the primary pulsation period. The Blazhko effect is not well understood, despite a number of investigations that are even based on unprecedentedly high-precision photometry from {\it Kepler} and {\it TESS} \citep{jurcsik2009, kolenberg2010, szabo2010, skara2020, molnar2022}. The space-based photometry has also revealed several low-amplitude additional modes and dynamical effects in the pulsation, including period doubling of the fundamental mode \citep{szabo2010, molnar2022}. While a significant fraction of RRL stars in GCs show modulations \citep{ferro2012, jurcsik2019}, we will refrain from a detailed discussion of the Blazhko effect in this article. 

\subsection{Color--Magnitude Diagrams}

\begin{figure}
\includegraphics[width=13.5 cm]{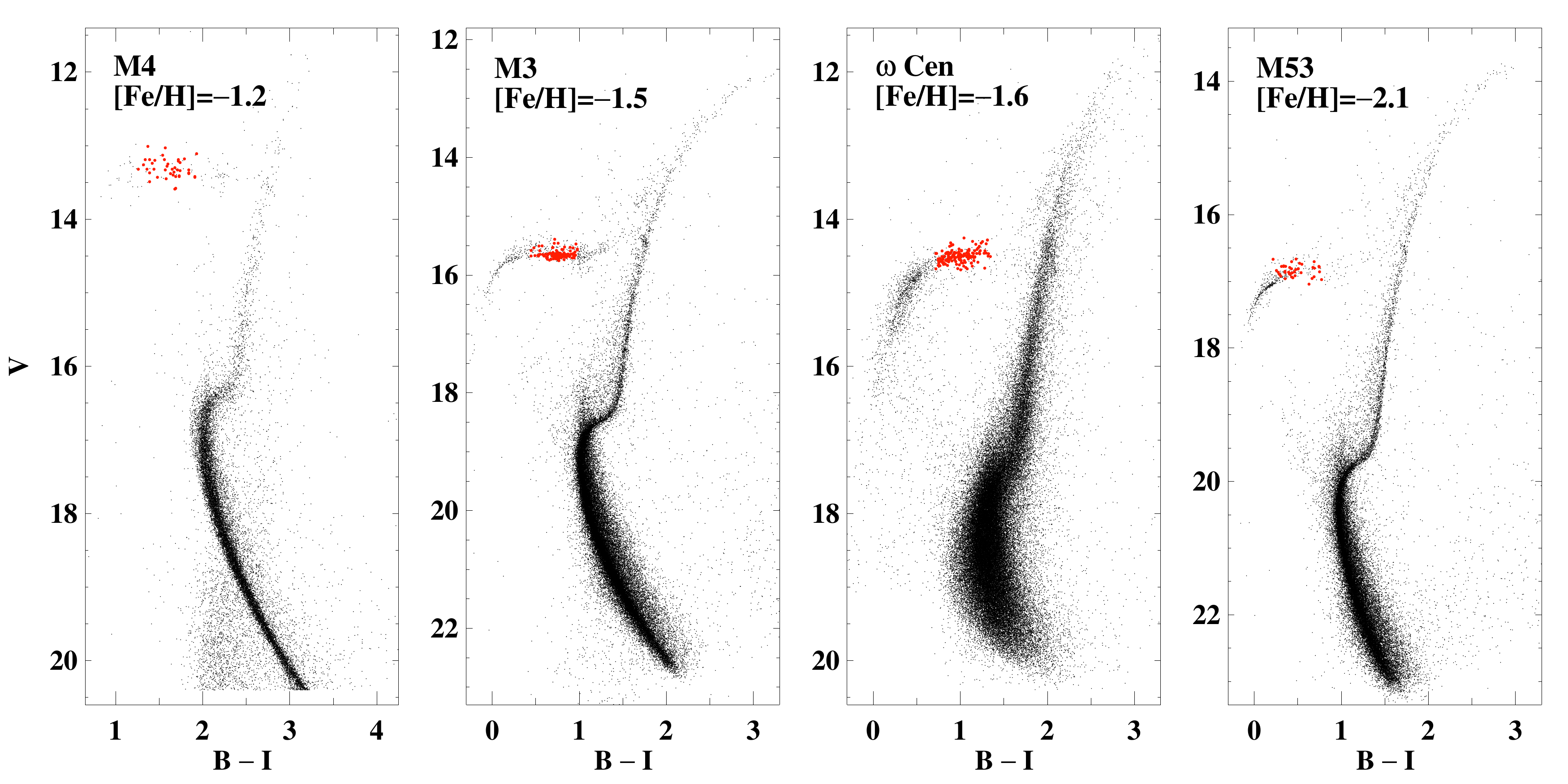}
\caption{Optical color--magnitude diagrams for GCs. RRL variables are shown in red symbols. \label{opt_cmd}}
\end{figure}

\begin{figure}[H]
\includegraphics[width=13.5 cm]{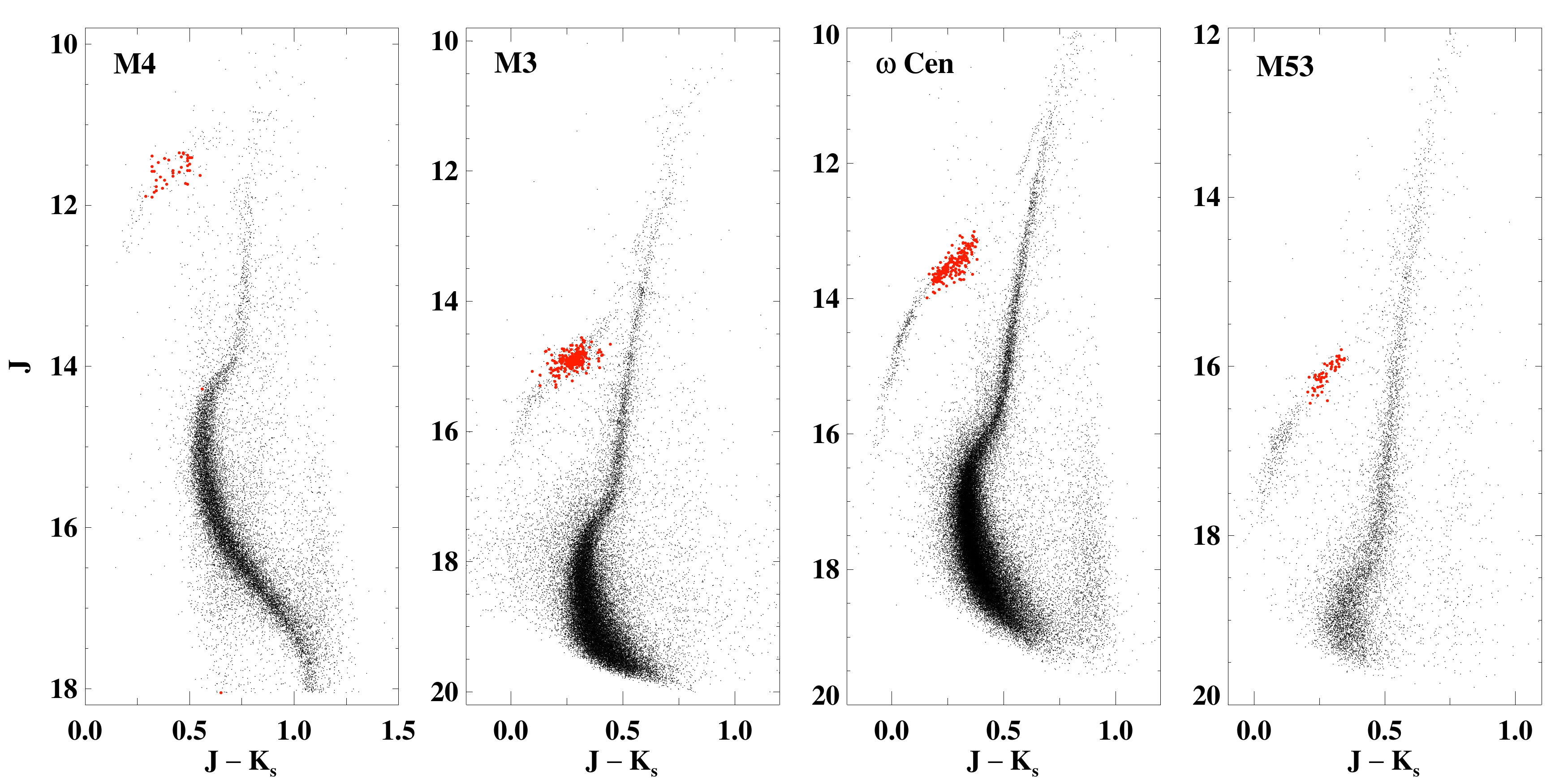}
\caption{NIR color--magnitude diagrams for GCs. RRL variables are shown in red symbols.\label{nir_cmd}}
\end{figure}

Figures~\ref{opt_cmd} and \ref{nir_cmd} show color--magnitude diagrams for M4 ([Fe/H]$\sim$$-$1.18), M3\linebreak ([Fe/H]$\sim$$-$1.50), $\omega$ Cen ([Fe/H]$\sim$$-$1.64), and M53 ([Fe/H]$\sim$$-$2.06) GCs in optical and NIR bands, respectively. The variable RRL population is also shown in colored symbols. These color--magnitude diagrams are not corrected for extinction; note that there is significant and differential reddening towards M4 and $\omega$ Cen \citep{hendricks2012, calamida2005}. 

The horizontal branch morphology is quite interesting and complex, and has been discussed extensively in several excellent reviews in the literature (\citep{smith1995, bono2003, catelan2009} and references therein). The main components of the HB include blue HB, variable HB (RRL gap), red HB, and the red clump \citep{catelan2009}. The red HB stars are commonly present in metal-rich and relatively young GCs. These occupy the horizontal part of the HB, unlike the more massive and significantly brighter red clump stars. Figure~\ref{opt_cmd} shows red HB stars in metal-rich M4 and M3, with the latter GC being more populated. A smaller fraction of red HB stars can also be found in the more metal-poor GCs, for example, as a result of the evolution of blue HB or RRL variables to the right of the color--magnitude diagrams. The variable HB that includes RRL variables is located in the region of the HB that crosses the Cepheid IS. In typical color--magnitude diagrams based on single-epoch data, this region appears as a ``gap'' because long-time monitoring is required to obtain the mean magnitudes of variable stars and to properly place those on the color--magnitude diagrams. However, the color--magnitude diagrams shown in Figures~\ref{opt_cmd} and in \ref{nir_cmd} are based on long-term photometric data, and therefore, the variable HB is distinctly evident. 

The blue HB consists of stars hotter than the RRL IS but is far from being horizontal at the blue end. The blue HB includes a canonical, ``horizontal'' part and a nearly ``vertical'' component which ends in a ``blue HB tail''. Figure~\ref{opt_cmd} and \ref{nir_cmd} show a distinctly clear blue HB for metal-poor GCs. The hottest stars on the HB have temperatures above 20,000~K, and are referred to as extreme or extended HB stars. These hotter HB stars are less luminous than the cooler blue HB stars and may also end up with a ``blue hook'' of extremely hot stars. Some studies have also suggested that the canonical blue HB and the blue HB tail are separated by a ``gap'' (see \citep{catelan2009} for details).

\begin{figure}[H]
\includegraphics[width=13.5 cm]{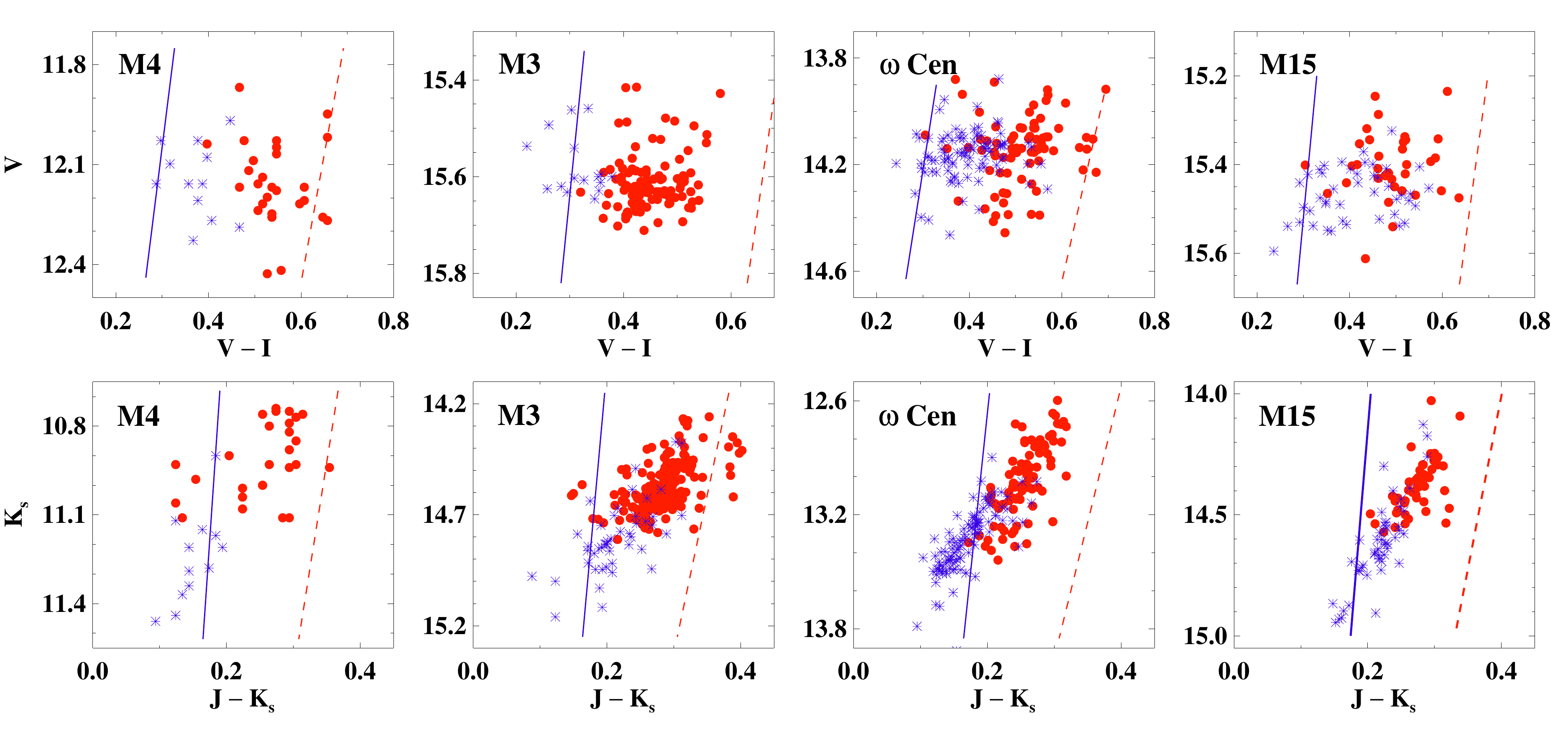}
\caption{Close-up of optical and NIR color--magnitude diagrams for RRL in different GCs. Red circles and blue asterisk show RRab and RRc, respectively. The theoretically predicted red and blue edges of the IS are also shown {from} 
\citet{marconi2015}.\label{rrl_cmd}}
\end{figure}

Figure~\ref{rrl_cmd} shows a zoom-in of the RRL color--magnitude diagrams in four different GCs. The magnitudes and colors are corrected for extinction using the classical reddening law of \citet{card1989}. Theoretically predicted boundaries of the fundamental red edge and the first-overtone blue edge of the IS are based on the empirical relations provided by \mbox{\citet{marconi2015}}. Theoretically predicted magnitudes were converted to apparent magnitudes using the distance to the GC in the respective studies \citep{braga2015, braga2018, bhardwaj2020, bhardwaj2021}. As expected, RRc are located towards the blue (hot) edge of the IS, while the RRab are towards the red (cool) edge of the IS. The red edge of the IS is consistent in both optical and NIR color--magnitude diagrams. The predicted blue edge appears to be redder in the case of M4 and $\omega$ Cen in NIR data, but is consistent in optical color--magnitude diagrams.
While same cluster reddening is adopted for all RRL, there is evidence of differential reddening in both these clusters. The agreement between predicted boundaries and empirical IS is reasonably good if we take into account the basic assumptions in pulsation models. The predicted boundaries of the IS were estimated for a fixed composition and the pulsation models also have a resolution of 50K in effective temperature. The differences can be partly attributed to the observed metallicity spread in $\omega$ Cen \citep{magurno2019}. Note that the predicted boundaries of the IS are metal-dependent in optical, but are derived to be independent of metallicity in NIR. 

\subsection{The Bailey Diagrams and the Oosterhoff Dichotomy}

The pioneering work of Bailey on variable stars in GCs initiated the investigations on period--amplitude diagrams for their RRL stars \citep{bailey1902}. The Bailey diagrams, or the period--amplitude diagrams for RRL differ from cluster to cluster, and these differences can be associated with the Oosterhoff type \citep{oosterhoff1939} of the GC. The Oosterhoff dichotomy is one of the long-standing problems in modern astronomy which shows that the galactic GCs can be separated into two distinct groups according to the mean period of their RRab variables. The RRab in Oosterhoff I (OoI) clusters have a shorter mean period of 0.55 days, and are relatively metal-rich ([Fe/H]~$\gtrsim$$-1.5$~dex), while RRab in Oosterhoff II (OoII) have a longer mean period of 0.65 days and [Fe/H]~$\lesssim$$-1.5$~dex \citep{oosterhoff1939, smith1995, catelan2009, catelan2015}.

\begin{figure}[H]
\includegraphics[width=13.5 cm]{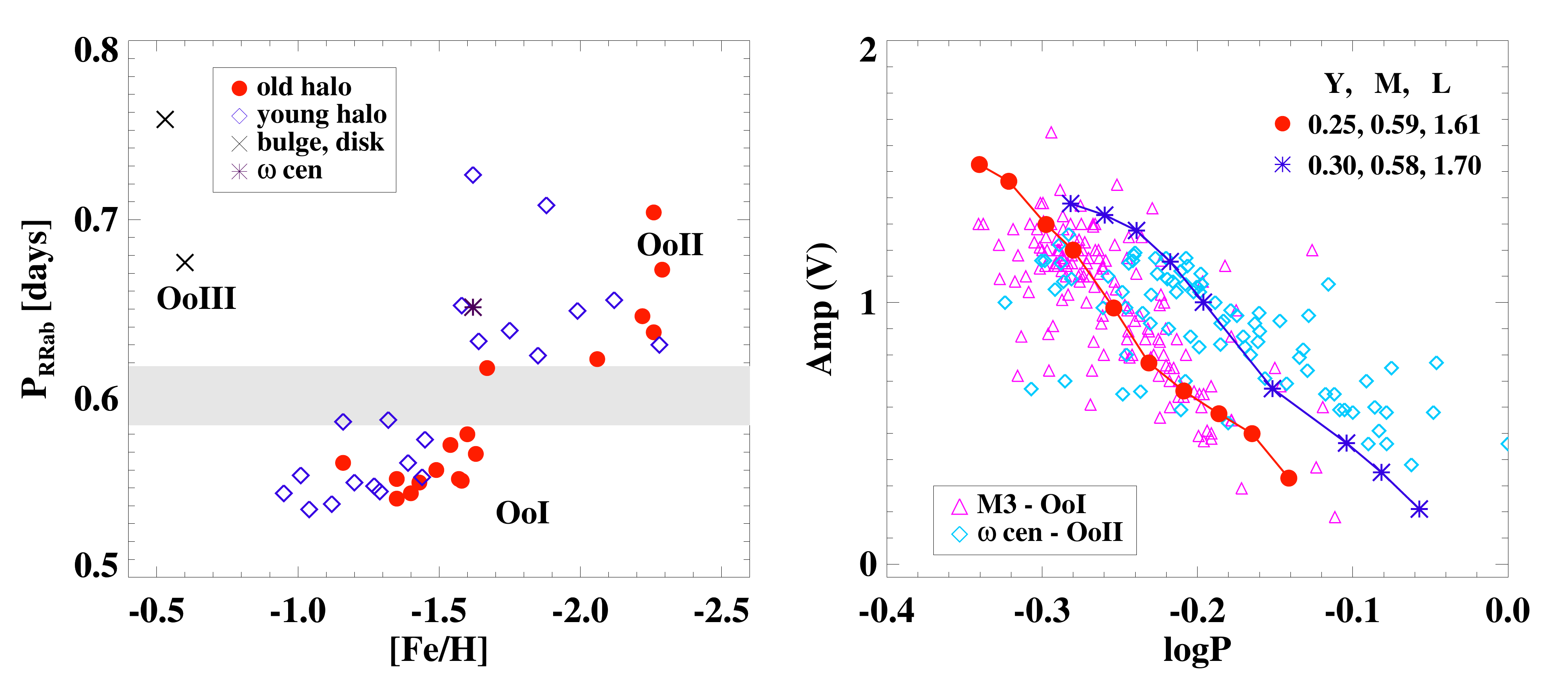}
\caption{{\textbf{Left:}} {Oosterhoff dichotomy} in GCs as shown by plotting the mean period of RRab stars versus [Fe/H] for Galactic GCs. The shaded region represents the Oosterhoff gap separating the two Oosterhoff types. Two metal-rich GCs (NGC 6441 and NGC 6388) are also shown as Oosterhoff III clusters. {\textbf{Right:}} Period--amplitude diagram for RRab in Oosterhoff I (M3) and II ($\omega$ Cen) clusters. For a fixed composition, RRL models {from} 
\citet{marconi2015, marconi2018} with different helium abundances, masses, and luminosities are also overplotted. \label{oo_pa}}
\end{figure}

The left panel of Figure~\ref{oo_pa} displays the Oosterhoff dichotomy in the Galactic GCs. These clusters are separated in two distinct groups of OoI and OoII, and a gap is seen in the mean period around 0.6 days. However, a few metal-rich clusters (NGC 6441, NGC 6388, [Fe/H] $>$$-$0.7 dex) have a longer value for the mean-period of RRab (0.7 days) and are often called OoIII clusters. Initially, the dichotomy was attributed to a dichotomy in the ``transition period'' between RRab and RRc stars, which reflects a difference in the effective temperature at the transition point \citep{vanalbada1973}. It was suggested that the temperature difference arising from the ``hysteresis effect'' \citep{vanalbada1973} is responsible for the Oosterhoff dichotomy. \mbox{\citet{sandage1958a}}, using a period--mean density equation, estimated that the $V$-band HB magnitudes differ by 0.2 mag between OoII and OoI clusters, such that the OoII clusters are brighter. \mbox{\citet{sandage1981}} quantified a period-shift at a fixed temperature and suggested that the Oosterhoff dichotomy can be explained as the difference in the intrinsic luminosity for the RRL in two clusters. \citet{renzini1983} suggested that, in addition to Sandage period-shifts, the non-monotonic behavior of HB morphology with decreasing metallicity can also contribute to the Oosterhoff dichotomy. Despite several investigations on Oosterhoff dichotomy, a general consensus has not yet been reached (see~\citep{catelan2009}). The GCs and dwarf galaxies in the Milky Way satellite systems do not show such a dichotomy (see \citep{catelan2009} for more details). This suggests that the Oosterhoff dichotomy is specific to galactic GCs and can provide useful insights into the Milky Way’s evolution history.

The right panel of Figure~\ref{oo_pa} shows a period--amplitude diagram of RRab variables in M3 and $\omega$ Cen. RRL in OoII clusters have longer periods than those in OoI clusters at a fixed amplitude. Most observational period--amplitude diagrams of RRab exhibit scatter in both OoI and OoII clusters. If the light curves are well-sampled, the photometric uncertainties on amplitudes are typically small (<0.05 mag). { The Blazhko effect in RRL, which can change mean amplitudes by more than $90\%$ (see Figure 4 of \citep{jurkovic2018}), is the dominant source of scatter in the observational period--amplitude diagrams.} In the case of $\omega$ Cen, the metallicity spread of RRL can also contribute to this scatter. RRL in OoI M3 also show a small fraction of OoII RRL~\citep{cacciari2005}, presumably because the mean metallicity of the GC is nearly at the boundary between OoI and OoII. The RRL models computed at fixed metal contents (\mbox{Z = 0.004}) and primordial helium contents of Y=0.25 and Y=0.30 are also shown in the left panel of Figure~\ref{oo_pa}~\citep{marconi2015, marconi2018, das2018}. The helium-enhancement results in a systematic shift in periods, which primarily occurs due to increased luminosity levels for similar masses (see \citep{rood1973, sweigart1998, marconi2018a} and references therein). Therefore, Bailey diagrams can be used to constrain the helium content of stars in a GC. \citet{marconi2018} estimated the helium-abundance (Y = 0.245) of the RRL population in the Milky Way bulge by comparing their minimum periods derived from stellar pulsation models. 

\begin{figure}[H]
\includegraphics[width=13.5 cm]{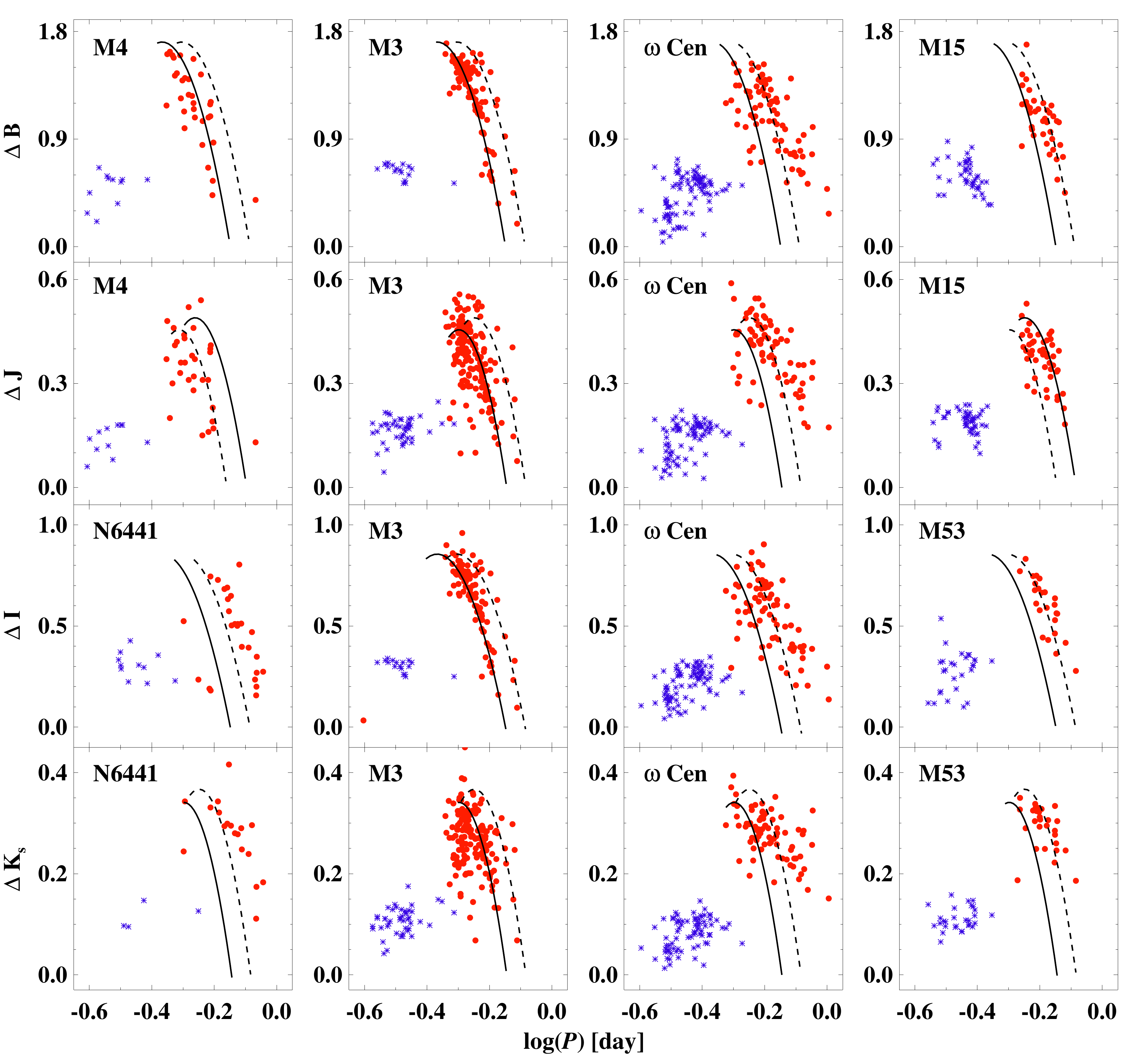}
\caption{{Optical} and NIR period--amplitude diagrams for RRL in GCs of different mean-metallicities. The solid and dashed lines represent loci of OoI and OoII RRab, respectively. \label{multi_pa}}
\end{figure}

Figure~\ref{multi_pa} shows optical and NIR period--amplitude diagrams for RRL stars in GCs of different mean-metallicities. The amplitudes of RRab stars decrease as a function of pulsation period, while the amplitudes for RRc seem to rise first and then fall with increasing period. The solid and dashed lines in the $B$- and $J$-band represent the approximate loci of the OoI and OoII RRab in M3 from \citet{cacciari2005} and \citet{bhardwaj2020a}, respectively. The loci in the $I$- and $K_s$-bands are overplotted by scaling the $B$- and $J$-band loci with corresponding amplitude ratios \citep{braga2018, bhardwaj2020a}. Most RRab in M4 fall on the locus of RRab stars while the majority of RRab in metal-poor M15 are located on the OoII locus. RRab in M3 seem to have both OoI and OoII sequences, with the former being clearly more populated. Infrared Bailey diagrams seem to exhibit larger scatter due to smaller amplitudes, which exhibit larger uncertainties because of the sparsely sampled light curves. In the case of $\omega$ Cen, there is more scatter, presumably resulting from intrinsic metallicity spread in RRL stars \citep{sollima2006a, magurno2019}. While the impact of metallicity on Bailey diagrams is debated \citep{bono2007}, period--amplitude--[Fe/H] relations have been used to determine metallicities for RRab stars (see, \citep{kinemuchi2006, kunder2009}). The period--amplitude--[Fe/H] relations suggest a continuous correlation between period and both the luminosity and metallicity for RRL in GCs.

In metal-rich OoIII NGC 6441, RRab do not fall clearly into either of the OoI or OoII groups in Figure~\ref{multi_pa}, but are closer to the OoII locus, similar to $\omega$ Cen. Even though NGC 6441 is metal-rich, previous studies of its RRab have shown that they have period--amplitude diagrams similar to those in OoII GCs \citep{corwin2006}. \citet{braga2020} showed that the long-period RRab in $\omega$ Cen follows the locus of OoIII RRab, similar to NGC 6441 variables. \mbox{\citet{rich1997}} first discovered that metal-rich NGC 6441 and NGC 6388 not only have expected red HB but also blue HB components that extend through the IS. These GCs are also suggested to host more than a single stellar population, with the possibly helium-enhanced smaller population responsible for the blue HB component \citep{mcwilliam2010, bellini2013}. In metal-poor clusters such as M15, there is considerable scatter in Bailey diagrams where a significant fraction of RRab fall between the two Oosterhoff-type sequences (e.g., \citep{corwin2008}). While there is some correlation between the location of Bailey diagrams and [Fe/H], several authors have suggested that the amplitude at a fixed period is a function of the Oosterhoff type, more than that of the metal abundance (e.g., \citep{bono2007, kunder2009}). The importance of Oosterhoff type becomes evident, for example, in OoI (M3) and OoII (M2) clusters, where the difference of mean-metallicities is small but M2 RRab are significantly shifted to longer periods than those in M3 (see Figure 4, p. 22 in~\citep{mcwilliam2010}).

Optical and NIR amplitude ratios have also been used as diagnostic tools for Oosterhoff dichotomy in RRab in GCs. \citet{braga2018} found that the optical--NIR and NIR amplitude ratios for long-period ($\log(P)>0.7$~days) RRab in $\omega$ Cen are systematically larger than the short-period RRab variables. \citet{bhardwaj2020a} studied optical and NIR amplitude ratios in M3 and found similar trends. The authors found a break period at ($\log(P)>0.6$~days), such that the amplitude ratios for the longer-period RRab are significantly larger. Median values of amplitude ratios are consistent between the two OoI and OoII clusters. The increase in amplitude ratios occur at a smaller period for RRab in M3 than those in $\omega$ Cen, and the difference in the break periods between the two clusters is similar to the difference in mean periods of their RRab stars. Therefore, amplitude ratios may also be indicators of the Oosterhoff type of the cluster. Furthermore, amplitude ratios are also important for template light curves of RRL, to correct for amplitudes when those are known only at a particular wavelength \citep{braga2019}.

\subsection{The Visual Magnitude--Metallicity Relation}

The absolute visual magnitudes of RRL stars have important astrophysical implications, in particular to determine distance using the magnitude--metallicity relation. \citet{sandage1981, sandage1982} correlated the location of RRL in the period--amplitude diagrams with both the absolute magnitude and [Fe/H], and determined the empirical relations between the period-shift and metallicity. From the observed period-shifts between clusters, \citet{sandage1982} derived an empirical relation between the luminosity and metallicity: $\Delta M_{bol}=0.35$[Fe/H]. The luminosity--metallicity relation between RRL $V$-band absolute magnitude ($M_V$) and the metallicity ([Fe/H]) is expressed as $M_V = \alpha + \beta\mathrm{[Fe/H]}$, where the slope ($\beta$) and the zero-point ($\alpha$) have been determined through several empirical and theoretical investigations in the literature (\citep{lee1990, fernley1998, caputo2000, clementini2003, bono2003, muraveva2018}, and references therein). The calibration of 
the magnitude--metallicity relation for RRL in the LMC \citep{clementini2003} and the Milky Way \citep{muraveva2018} are provided below:
\begin{align}
\nonumber
V_0 &= 19.064(\pm0.017) + 0.214(\pm0.047)\mathrm{[Fe/H]}, \nonumber\\ 
M_V &= 1.17(\pm0.04) + 0.34(\pm0.03)\mathrm{[Fe/H]}. \nonumber
\end{align}

\textls[-15]{The zero-point in the first relation above includes the distance to LMC in the case of apparent magnitude ($V_0$). The absolute calibration ($M_V$) has been derived by \mbox{\citet{muraveva2018}}} using {\it Gaia} parallaxes for field RRL. While the magnitude--metallicity relation is simple and straightforward in the sense that it requires only two observables---the apparent visual magnitude and the metallicity---its calibration is affected by deceptive systematic errors, which leads to a range of metallicity coefficients (slopes) from 0.37 to \mbox{0.13 mag dex$^{-1}$~\citep{muraveva2018}}. Several studies have also provided theoretical and empirical evidences that the $M_V-$[Fe/H] relation is not linear over the entire metallicity range covered by the GCs, and proposed a quadratic relation \citep{caputo2000, bono2003, catelan2004, bono2007}. The shallower slope for metal-poor systems becomes significantly steeper for metal-rich ([Fe/H] $>$$-$1.5) RRL stars. {However, this departure from linearity is not observed for LMC RRL stars over a broad metallicity range \citep{clementini2003}, which indicates the difficulty in disentangling the metallicity and evolutionary effects in the observational magnitude--metallicity relation.} With Gaia parallaxes, \mbox{\citet{muraveva2018}} found that the coefficient of the quadratic metallicity term is not significant, and that the zero-points are consistent between both linear and quadratic versions. 

The magnitude--metallicity relation relies on the assumption that RRL stars are on the ZAHB, but both the field and cluster RRL exhibit a spread in luminosity. Therefore, the theoretical calibrations based on ZAHB models do not account for evolutionary effects~\mbox{\citep{catelan2004, bono2007}}, while the empirical calibrations use both ZAHB RRL and evolved RRL where the latter have higher luminosities \citep{beaton2018}. The intrinsic luminosity width of the ZAHB becomes larger when moving from metal-poor to metal-rich GCs. Therefore, RRL with different metal abundances are affected by the evolutionary effects of both off-ZAHB evolution and the HB morphology \citep{caputo1998}. One of the sources of systematics on empirical magnitude--metallicity relations is the reddening uncertainties which are significant, i.e., a small uncertainty of 0.01 mag in reddening values implies an error of 0.03 mag in the visual magnitude. Furthermore, the variation in the initial helium content, which may depend on environment, can also affect the visual magnitudes of RRL variables \citep{beaton2018}. 

\subsection{Multiband Period--Luminosity Relations}

The pulsation theory predicts a period--luminosity--color relation for classical pulsating stars such as RRL and Cepheid variables. While Cepheids follow well-defined PLRs at both visual and infrared wavelengths, the dependence of RRL luminosity on pulsation period is negligible at visual wavelengths. Unlike Cepheids, which cover a wide range of luminosity within the IS, RRL have much smaller variation in the luminosity within the HB phase of low-mass stars. Given that the luminosity variations for RRL are much more modest, the temperature range within the IS also plays an important role in determining the range of RRL periods. Therefore, RRL stars exhibit a PLR only if the bolometric corrections are significant and are such that the absolute magnitudes cover a large range, as is the case at longer wavelengths. Indeed, the bolometric correction’s sensitivity to effective temperature becomes signiﬁcant only at wavelengths longer than the $R$-band \citep{catelan2004, marconi2015}, resulting in a well-defined PLR. 

\begin{figure}[H]
\includegraphics[width=13.5 cm]{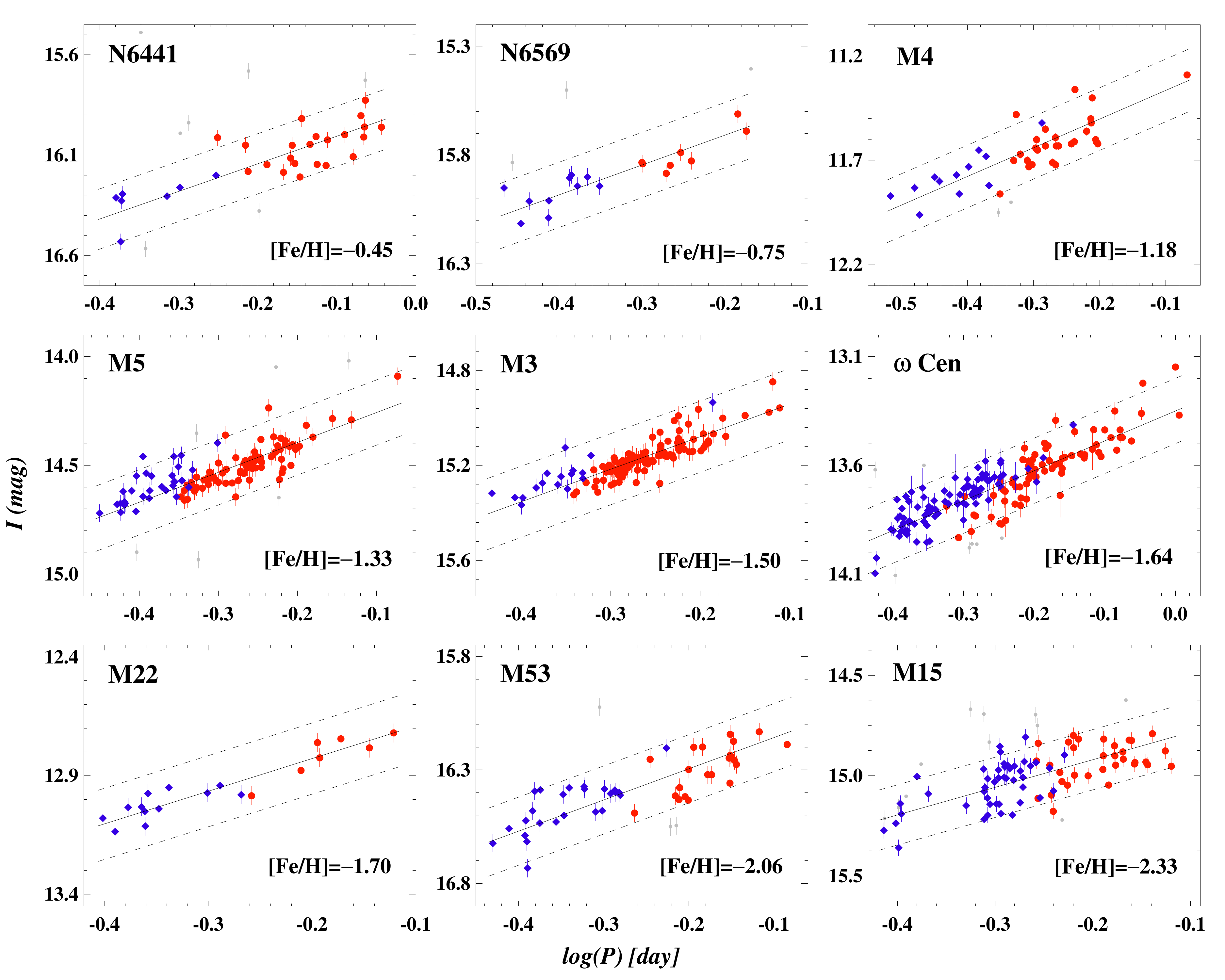}
\caption{$I${-band PLRs for RRL }in GCs of different mean-metallicities. Solid lines display a universal slope of PLR relation to all GCs while dashed lines represent $\pm3\sigma$ scatter in the underlying relation. Mean-metallicities for each cluster are also indicated in each panel. Red circles and blue squares represent RRab and (fundamentalized) RRc stars, respectively. Grey symbols represent $\pm3\sigma$ outliers. \label{opt_plr}}
\end{figure}

The pioneering work of \citet{longmore1986} showed that RRL stars follow a very tight PLR in the $K$-band. Using single-epoch $K$-band photometry in three GCs (M3, M5, and M107), \citet{longmore1986} found that the PLRs exhibit remarkably small scatter as compared to optical bands. The infrared PLRs come naturally from the pulsation equation because bolometric corrections increase with effective temperature, such that redder RRL are brighter in the $K$-band. NIR observations of RRL in the earliest studies on PLRs were limited to a few epochs or sparsely sampled light curves (e.g., \citep{longmore1990, nemec1994, butler2003}). With the advent of modern infrared detectors, several empirical investigations have been carried out on infrared PLRs for RRL stars in GCs (\citep{coppola2011, braga2015, neeley2017, braga2018, bhardwaj2020, bhardwaj2021}, and references therein). 

We compared PLRs for RRL in GCs with different mean-metallicities for which time-series optical and NIR data is available in the literature. The time-series photometry is critical to obtain precise and accurate mean-magnitudes for well-constrained PLRs. Figure~\ref{opt_plr} shows $I$-band PLR for RRL in GCs covering a wide metallicity range of $\Delta$[Fe/H]$\sim$2.0~dex. Under the assumption of the universality of RRL PLRs in GCs, we solved for a common slope and variable zero-point in different GCs. Note that the periods of overtone-mode RRc variables were fundamentalized using $\log(P_{RRab}) = \log(P_{RRc}) + 0.127$ \citep{petersen1991,coppola2015} to obtain a global (RRab+RRc) sample of RRL. Outliers beyond $\pm3\sigma$ were removed recursively until convergence. Optical mean-magnitudes are based on long-term photometric light curves and are, therefore, very accurate and precise. However, there are possible systematics in the data adopted from different literature studies. Nevertheless, a common slope of the PLR fits the data for all GCs very well, regardless of their metallicity, thus validating the assumption of universality of RRL PLRs.

\begin{figure}[H]
\includegraphics[width=13.5 cm]{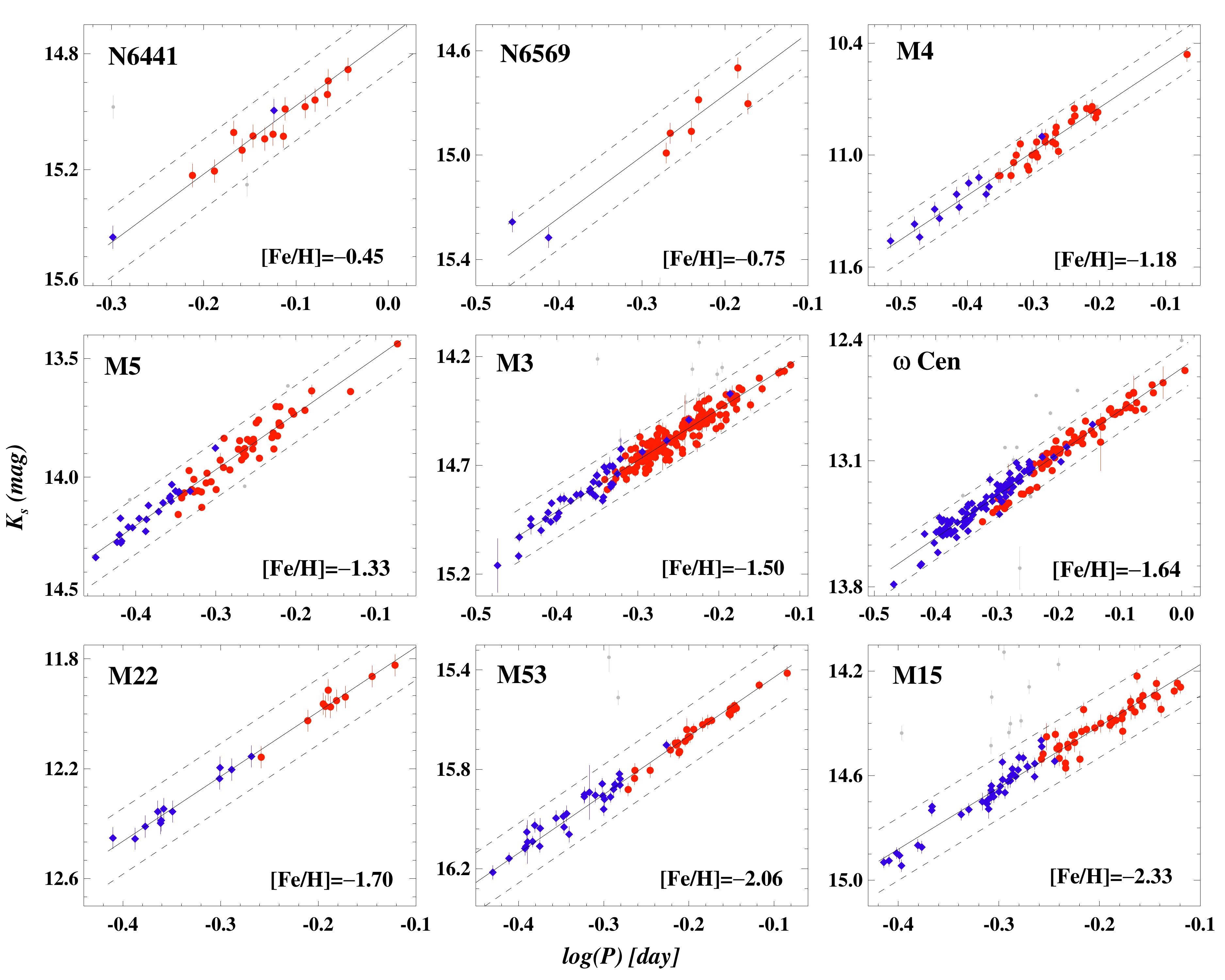}
\caption{$K_s${-band PLRs for} RRL in GCs of different mean-metallicities. Solid lines display a universal slope of PLR relation to all GCs, while dashed lines represent $\pm3\sigma$ scatter in the underlying relation. Mean-metallicities for each cluster are also indicated in each panel. Red circles and blue squares represent RRab and (fundamentalized) RRc stars, respectively. Grey symbols represent $\pm3\sigma$ outliers. \label{ir_plr}}
\end{figure}

NIR photometric light curves are much more sparsely sampled than the optical light curves (see Figure~\ref{rrl_lcs}). Therefore, NIR templates for RRL become very useful for determining accurate mean-magnitudes and deriving precise PLRs. Given that the NIR light curves are much more symmetric and near-sinusoidal in shape than the optical light curves, a simple sine-series fit is the simplest template that can be adopted to estimate mean-magnitudes and peak-to-peak amplitudes. While sine series is a good approximation for RRc light curves, the RRab light curves exhibit clear evidence of asymmetry for different periods (see Figure~\ref{rrl_lcs}). \citet{hajdu2018} used principal component analysis to generate $J$ and $H$-band templates from well-sampled $K_s$-band light curves of RRL in the VISTA VVV survey \citep{minnitivvv2010}. Similarly, \citet{braga2019} provided new $JHK_s$ template light curves of RRL variables and showed that the mean magnitudes can be estimated with an accuracy of $2\%$, even from single-epoch NIR observations. Figure~\ref{ir_plr} displays $K_s$-band PLRs for RRL in different GCs. The mean-magnitudes for RRL in these GCs are based on template-fits to $K_s$-band light curves. The significantly smaller scatter, as compared to the optical bands, is distinctly evident in the PLRs. Even in the $\omega$ Cen, which exhibits a spread in metallicity, the $K_s$-band PLR does not exhibit large intrinsic dispersion \citep{navarrete2015, braga2018}.

Similar to NIR bands, mid-infrared (MIR) observations also have indisputable advantages with respect to shorter wavelengths because the extinction is more than an order of magnitude smaller ($A_V$$\sim$$ 15A_{3.6\upmu \rm m}$) at the 3.6-$\upmu$m band. The luminosity variations are also small and mostly insensitive to effective temperatures. A few studies utilized increasing MIR observations, in particular, from the Wide Field Infrared Survey Explorer (WISE) and the {{Spitzer Space Telescope} 
} to investigate MIR PLRs for field RRL \citep{klein2011, madore2013, klein2014} and cluster RRL~\citep{dambis2014, neeley2017, muraveva2018a}. The calibration of MIR PLRs of RRL in 15 GCs was first provided by \citet{dambis2014} using WISE data. Depending on the calibration based on either statistical or trigonometric parallaxes, \citet{dambis2014} found two significantly different estimates of the absolute zero-point of PLRs. Together with a theoretical framework, \mbox{\citet{neeley2017}} employed {{Spitzer}} observations of RRL in M4 to derive precise PLRs with a dispersion of $\sim0.05$ mag in 3.6-$\upmu$m and 4.5-$\upmu$m bands as part of the {\it {Carnegie RR Lyrae program}}. The authors showed that the metallicity spread introduces a dispersion of 0.13 mag in the PLR if the sample includes a range of iron abundances. However, the scatter in MIR PLRs is reduced to 0.02 mag when the metallicity term is included, implying a percent-level precise distance determination using RRL stars.

\subsection{Period--Luminosity--Metallicity Relations}

Theoretical studies on RRL predicted a strong dependence of PLRs on metal abundance at all wavelengths. \citet{bono2001} derived a theoretical $K$-band PLZ relation and showed that the dependence of RRL luminosity on metallicity ($\sim$0.17 mag/dex) is relatively smaller than that in the optical bands (>0.2 mag/dex). Additionally, the uncertainties on the mass and luminosity also do not effect the PLRs significantly at infrared wavelengths. However, \citet{sollima2006} derived the empirical period--luminosity--metallicity (PLZ) relation in the $K_s$-band for RRL in GCs and found a metallicity coefficient ($\sim$0.08 mag/dex) nearly two-times smaller than the theoretical predictions of \citet{bono2001}. \citet{catelan2004} also derived metal-dependent PLRs for RRL based on synthetic HB models and found a significant metallicity term (0.21--0.17 mag/dex) in $IJHK_s$ bands. Using stellar pulsation models, \citet{marconi2015} generated RRL models for a wide range of metal abundances (Z~=~0.02 to 0.0001) at $UBVRIJHK_s$ wavelengths. They also found a strong metallicity dependence that varies from $\sim$0.14--0.19 mag/dex depending on the subclass of RRL and the wavelength under consideration. Recently, \citet{bhardwaj2021} investigated the $K_s$-band PLZ relation for RRL in GCs and found a metallicity coefficient consistent with theoretical predictions. The PLZ relations for RRL are written in the following form:
\begin{equation}
\label{eq:rrl_plz}
M = \alpha + \beta\log(P) + \gamma[Fe/H],\\
\end{equation}

\noindent where $M$ is the absolute magnitude, and $\alpha$, $\beta$, and $\gamma$ are the slope, zero-point, and the metallicity coefficient, respectively.

\begin{figure}[H]

\includegraphics[width=8.5 cm]{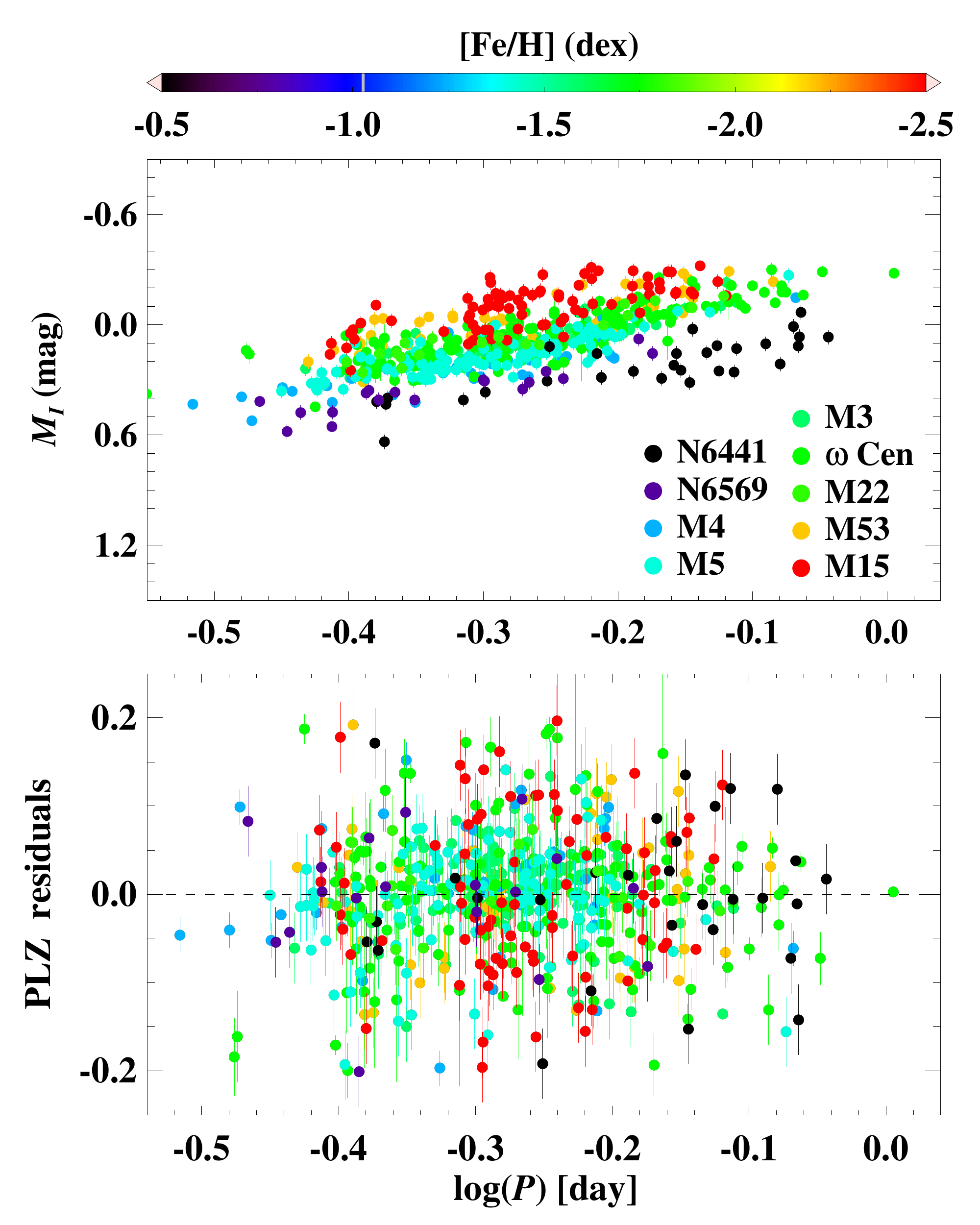}
\caption{{\bf Top:} {Combined} $I$-band PLRs for RRL in different GCs. The absolute magnitudes for RRL in GCs were determined based on the zero-point calibration using {\it Gaia EDR3} parallaxes. The color code in the legend corresponds to the mean metallicity of GC. 
{\bf Bottom:} residuals of the PLZ relation in the $K_s$-band for RRL stars. Note the strong metallicity trend in the PLRs in the top panel, which disappears in the residuals of the PLZ relation. \label{opt_plz}}
\end{figure}

Figure~\ref{opt_plz} displays the $I$-band PLZ relation for RRL in GCs. We used the mean-magnitudes derived from the time-series data to obtain an accurate and precise relative quantification of the metallicity coefficient. The mean-metallicity for these GCs is taken from \citet{carretta2009} for consistency with the theoretical models of \citet{marconi2015}. Note that one main source of systematics is the independent distances to GCs. As shown in Figure~\ref{opt_plr}, we can solve for a universal slope and separate the zero-point for each GC simultaneously. For the absolute quantification of metallicity effects and the zero-point, we adopted theoretical absolute magnitudes and the absolute magnitudes based on parallaxes of field RRL from {\it Gaia EDR3} \citep{lindegren2021}, as discussed in detail in \citet{bhardwaj2021}. The empirical PLZ relation in the $I$-band is shown in Figure~\ref{opt_plz}. The top panel displays clear trend in RRL magnitudes as a function of metallicity, such that the metal-poor RRL are brighter. When the metallicity term is taken into account, the residuals of the PLZ relation, as shown in the bottom panel of Figure~\ref{opt_plz}, do not exhibit any trend with the metallicity.

The empirical PLZ relation in the $I$-band derived in this work and the theoretical PLZ relation from \citet{marconi2015} are provided below:

\begin{align}
\label{eq:rrl_plr_model}
\textrm{M}_{I_{GC}} &= 0.05 -1.38\log(P) + 0.21\textrm{[Fe/H]} ~~~(\sigma=0.07), \nonumber\\ 
\textrm{M}_{I_{TH}} &= -0.07 -1.53\log(P) + 0.17\textrm{[Fe/H]} ~~~(\sigma=0.09).
\end{align}

Here, the uncertainties on the coefficients are small, and therefore, the difference in the slope of the empirical and predicted PLZ relation is statistically significant. The slope of the empirical PLZ relation is shallower than the theoretical relations, and the metallicity coefficients are relatively larger than the theoretical predictions. This difference in the slope of the empirical PLZ relation could possibly be due to the bulge GCs, which seem to show shallower PLR than the more metal-poor clusters. Note that the extinction in the bulge clusters is large at optical wavelengths, and the reddening law also deviates from the standard \citet{card1989} law. Moreover, the optical data is taken from several different studies in the literature, and there could be possible systematic uncertainties due to the inhomogeneous sample of optical photometry in GCs. 

\begin{figure}[H]

\includegraphics[width=8.5 cm]{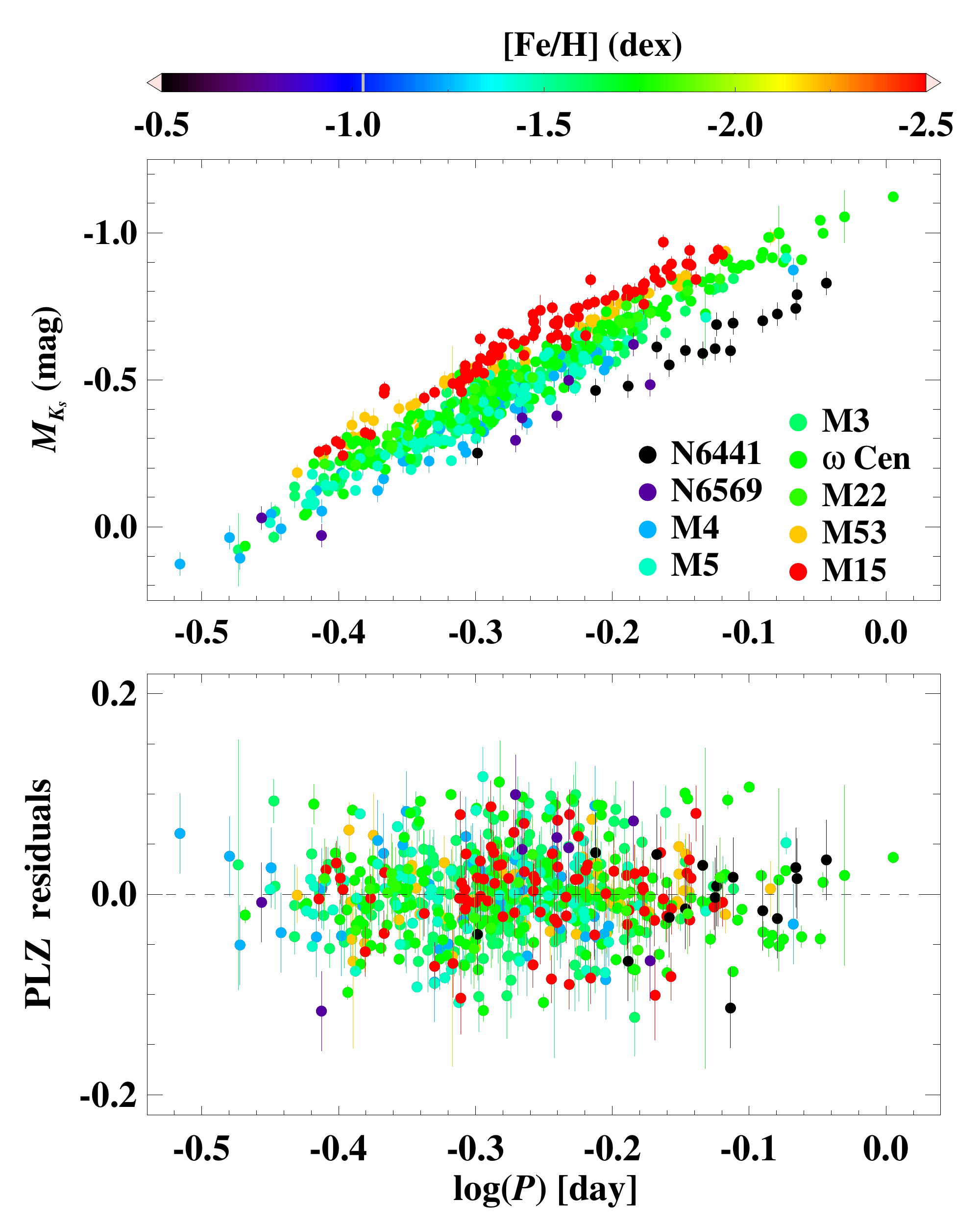}
\caption{Same as in Figure~\ref{opt_plz}, but in the $K_s$-band. \label{ir_plz}}
\end{figure}

Figure~\ref{ir_plz} displays the $K_s$-band PLZ relation for RRL in GCs. The empirical relation, together with the theoretical $K$-band relation from \citet{marconi2015}, is:
\begin{align}
\label{eq:rrl_plr_model}
\textrm{M}_{K_{GC}} &= -0.86 -2.35\log(P) + 0.17\textrm{[Fe/H]} ~~~(\sigma=0.04), \nonumber\\ 
\textrm{M}_{K_{TH}} &= -0.82 -2.25\log(P) + 0.18\textrm{[Fe/H]} ~~~(\sigma=0.04).
\end{align}

The uncertainties in the coefficients of the $K_s$-band PLZ relations are marginal, given the tight relation ($\sigma \sim$0.04~mag) for RRL in the GCs. Note that all $K_s$-band photometry is homogenized in the 2MASS photometric system. The empirical and theoretical calibrations of PLZ relation are in excellent agreement in the $K_s$-band. The absolute calibration of the RRL PLZ relation using {\it Gaia} EDR3 parallaxes also results in a zero-point consistent with theoretical relations. While we did not discuss the application of NIR PLZ relations for RRL to individual stellar systems, several investigations have been carried out in the literature to obtain precise and accurate distances to GCs, e.g., M92 \citep{del2005}, M5 \citep{coppola2011}, $\omega$ cen \citep{navarrete2015, braga2018}, M4~\citep{braga2015}, M3 \citep{bhardwaj2020a}, and M53 \citep{bhardwaj2021}. 

In addition to PLZs for RRL, Wesenheit magnitudes \citep{madore1982, bhardwaj2016a} have also been used to derive period--Wesenheit--metallicity (PWZ) relations. Wesenheit magnitudes are, by construction, independent of reddening, and therefore, two-color photometry allows for accurate distance determinations using PWZ relations by minimizing extinction uncertainties. \citet{marconi2015} presented theoretical optical and NIR PWZ relations using stellar pulsation models. The authors found that the optical Wesenheit, using a combination of $B$- and $V$-bands, is almost independent of metallicity, but other Wesenheit relations exhibit relatively strong metallicity dependence. The application of PWZ relations to obtain a precise distance to $\omega$ Cen has been discussed in detail in \citet{braga2016}. 

{ The absolute calibration of RRL PLZ and PWZ relations is critical for providing classical Cepheid-independent calibrations of the extragalactic distance scale for the determination of the Hubble constant \citep{beaton2018}. The Hubble constant measurements in the late and early Universe are currently in tension, implying that the Universe is expanding faster than the predictions of the standard cosmological model (see \citep{verde2019, riess2021, freedman2021} for details). This Hubble tension, if confirmed, could have significant consequences for both fundamental physics and modern cosmology. Therefore, independent extragalactic distance scales based on population I and population II distance indicators are crucial for exploring residual systematics in the local Hubble constant measurements. RRL and the Tip of the Red Giant Branch (TRGB) stars are population II distance indicators that can be applied to galaxies of any Hubble type and of any inclination, and they are found in less-extincted and less-crowded environments, unlike young population I Cepheids \citep{freedman2020}. Given the recent progress in observational and theoretical frameworks, the stage is set for fully exploiting population II distance indicators. RRL, despite being fainter than Cepheids and TRGB, can be used to complement and test TRGB distances within and beyond the local group. The GC RRL variables can be used to determine their accurate distances and calibrate independent secondary distance indicators such as the GC luminosity function. The improvement of RRL as distance indicators will not only enable morphological studies of the Milky Way and nearby galaxies, but also boost their application as stellar population tracers of age, metallicity, and the extinction of their host galaxies.
The RRL distance scale will reach its full potential with upcoming major observational facilities, which will predominantly operate in the infrared regime where these standard candles become more precise and powerful.}

\section{Type II Cepheids}
\label{sec:sec6}

The discovery of T2Cs was a crucial step in the modernization of the extragalactic distance scale in the 1950s, when it was suggested that the PLRs of T2Cs are different than those of classical Cepheids \citep{baade1958}. Since then, T2Cs have received far less attention than classical Cepheids because the former are fainter and typically less abundant. In the catalogue of \citet{clement2001}, there are about hundred known candidate T2Cs in the GCs. More than 300 T2Cs are known in the Magellanic Clouds, but the number is nearly thirty-times smaller than that of known classical Cepheids \citep{soszynski2018, soszynski2015}. However, T2Cs are very useful tracers in old stellar systems such as the Galactic bulge \citep{soszynski2017}, where population I classical Cepheids are not present. \citet{soszynski2017} found more than 900 T2Cs in the bulge. 
T2Cs are known to have evolved from the blue HB in GCs, and therefore, they are likely to be found in stellar systems where blue HB is well populated. 

Most GCs host, at most, a couple of T2Cs; thus, homogeneous optical and NIR photometry is not available for a statistically significant number of cluster variables in the literature. Only three clusters (NGC 6388, NGC 6441, and $\omega$ Cen) host seven or more T2Cs, with NGC 6388 known to host 10 T2C candidates \citep{clement2001}. \citet{nemec1994} provided a compilation of optical photometry of 40 GC T2Cs and derived their PLRs. \citet{matsunaga2006} used homogeneous NIR time-series for 46 T2Cs in 26 GCs to derive $JHK_s$ PLRs. While \citet{pritzl2003} added T2Cs in NGC 6388 and NGC 6441 to an optical sample, \citet{braga2020} provided an updated sample of T2Cs with NIR photometry by adding variables in $\omega$ Cen. We combined and updated the optical and NIR dataset of T2Cs from these studies to obtain a sample of 60 variables in GCs for a comparative study of their pulsation properties. The distances, mean-metallicities, and the reddening values in each GC were also compiled from the aforementioned studies. The mean-magnitudes and colors of T2Cs were also corrected for extinction using the reddening law of \citet{card1989}.

\subsection{Multiband Light Curves}

\begin{figure}[H]
\includegraphics[width=13.5 cm]{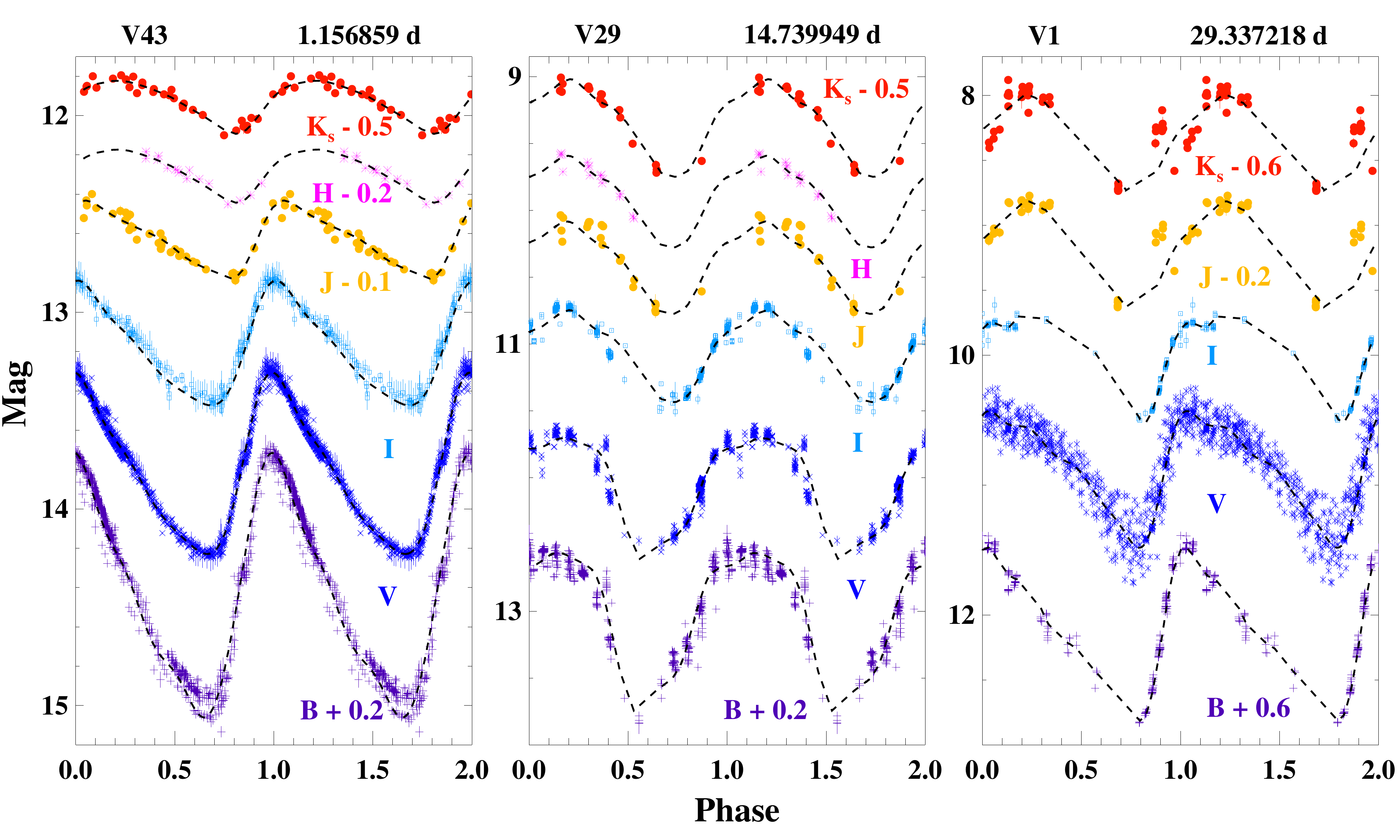}
\caption{Multiband light curves of T2Cs in $\omega$ Cen {from} 
\citet{braga2020}. Each panel displays example light curves of different subclasses of T2Cs---{\bf Left:} BL Her; {\bf Middle:} W Vir; {\bf Right:} RV Tau. The variable star ID and the pulsation periods are listed on the top of each panel. \label{lcs_t2c}}
\end{figure}

T2Cs are known to exhibit complex light curve variations depending on their subtypes, periods, and stellar environment. Figure~\ref{lcs_t2c} displays example light curves of a BL Her, W Vir, and RV Tau variable in $\omega$ Cen. A clear decrease in amplitude and a small phase lag from optical to NIR is evident in these light curves. The optical light curves of BL Her show a sharp rise from minima to maxima and a shallower decreasing branch similar to RRab stars (see Figure~\ref{rrl_lcs}). This saw-tooth RRab-like characteristic is associated with the above HB variables originally suggested by \citep{diethelm1983} for short-period (<3~days) T2Cs. NIR light curves are also more symmetric in $HK_s$-bands. T2Cs in the bulge display secondary bumps on the descending branch for the shortest periods, which moves to earlier phases, and to the ascending branch for longer periods \citep{sandage1994, soszynski2018}. Note that this is similar to the so-called ``Hertzsprung Progression'' in classical Cepheids \citep{hertzsprung1926, bhardwaj2015, bhardwaj2017}. This bump feature disappears for longer-period ($P>3$~days) BL Her, which show a plateau or flat maxima (e.g., V48 in Figure B.1 of \citep{braga2020}). 

The multiband light curves of a W Vir T2C, shown in Figure~\ref{lcs_t2c}, are near-sinusoidal, in particular for wavelengths longer than $V$. The variations in the light curves occur systematically, which can change amplitudes in different pulsation cycles. For shorter-period ($5<P<8$~days) W Vir, the rise time from minima to maxima is faster, but the light curves are more symmetric for longer-period variables. For longer-period W Vir in the bulge and Magellanic Clouds, a bump prior to the maxima is commonly seen, as first noticed by \citet{diethelm1983}. Another class of peculiar W Vir was also proposed by~\mbox{\citet{soszynski2008}} in the Magellanic Clouds, which are brighter and bluer than W Vir. The light curves of peculiar W Vir exhibit significant scatter due to variations of periods and additional modulations, suggesting their binary evolution. 

The light curves of RV Tau are known to exhibit alternating deep and shallow minima. However, the light curves of V1, shown in the right panel of Figure~\ref{lcs_t2c}, do not display this feature, much like the other RV Tau variables in GCs \citep{zsoldos1998}. The alternative minima is typical of field RV Tauri, and is also common in bulge and Magellanic Cloud variables. The occurrence of deep and shallow minima also interchanges in some RV Tau variables. Furthermore, the maxima can also exhibit alternations such that higher maxima follow deeper minima. The light curves of RV Tau exhibit significant scatter due to their irregular brightness variations from cycle to cycle, and this scatter increases for longer-period RV Tau stars. 

\subsection{Color--Magnitude and Period--Amplitude Diagrams}

\begin{figure}[H]

\includegraphics[width=10.5 cm]{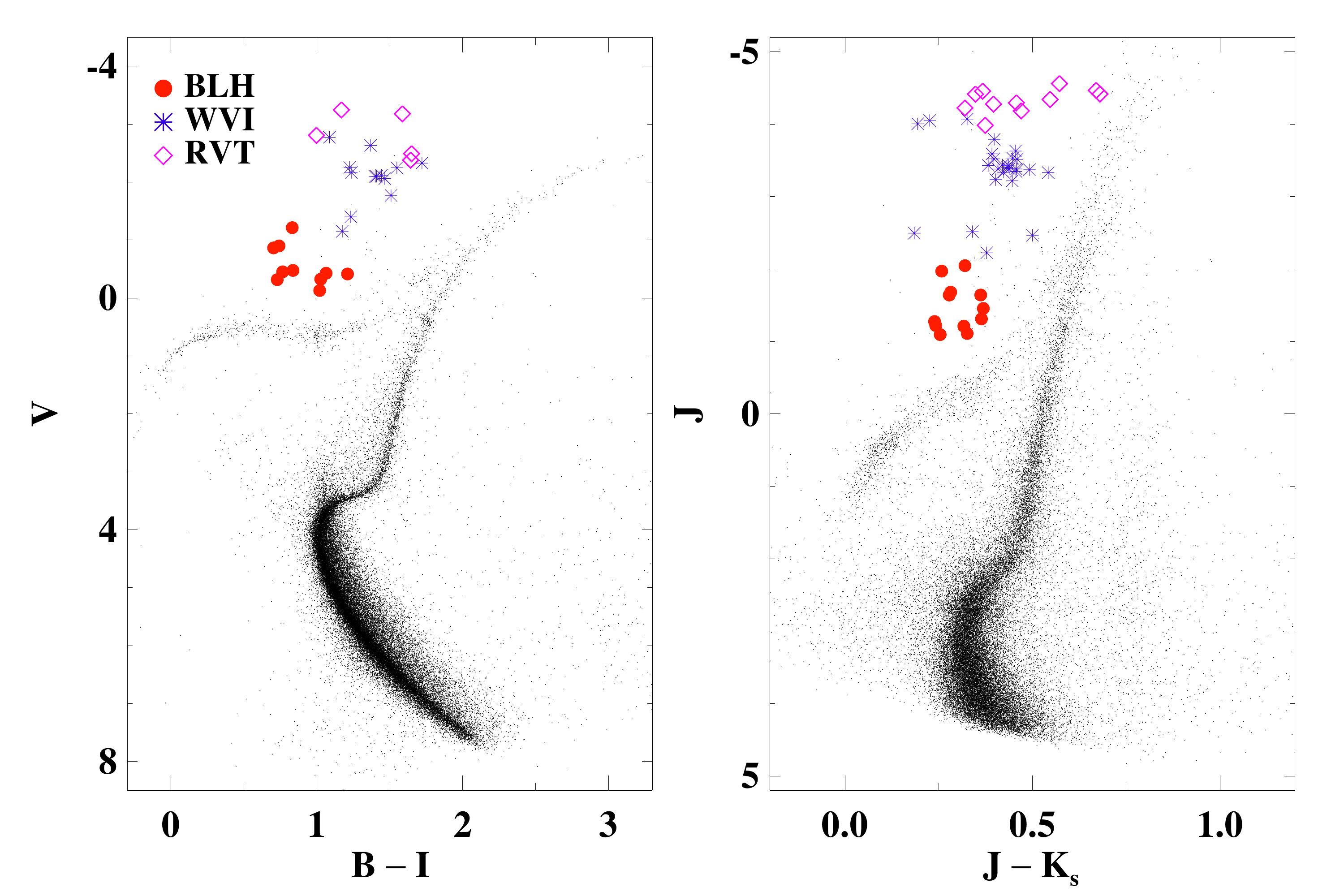}

\caption{Optical and NIR color--magnitude diagrams of different subtypes of T2Cs variables. BLH---BL Her, WVI---W Vir, and RVT---RV Tau. The reference color--magnitude diagram (small/black data points) is of M3, after correcting for distance and extinction. \label{t2c_cmd}}
\end{figure}

Figure~\ref{t2c_cmd} displays optical and NIR color--magnitude diagrams of T2Cs after correcting for distance and extinction. The short-period BL Her are slightly bluer than the longer period T2Cs, as these evolve from the blue HB towards the AGB and cross the IS while depleting helium in the core. The brighter W Vir are considered to make temporary excursions through IS from cooler temperatures, when they suffer shell flashes after reaching AGB \citep{wallerstein2002}. However, these excursions are not expected to occur commonly, as suggested by the theoretical models \citep{smolec2016, bono2016}.
Note that there are fewer T2Cs in optical color--magnitude diagrams because our sample lacks quality data for long-period RV Tau. Long-term photometric monitoring is essential to properly estimate the mean pulsation properties for these variables. 

\begin{figure}

\includegraphics[width=13 cm]{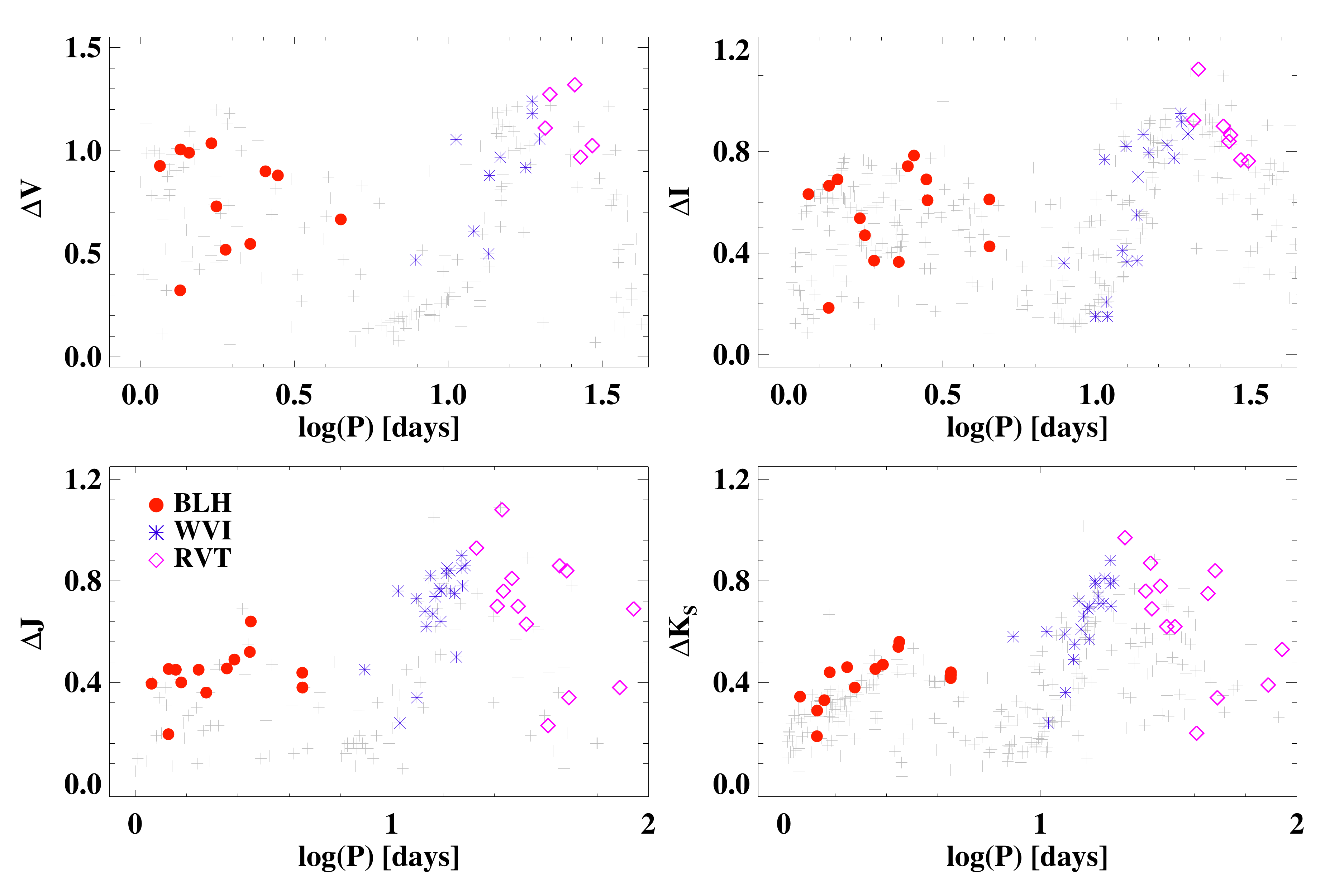}

\caption{Optical and NIR period--amplitude diagrams for T2Cs in GCs. The grey symbols are amplitudes of T2Cs in the LMC (\textbf{left} panels) and the Galactic bulge (\textbf{right} panels). \label{t2c_amp}}
\end{figure}

Figure~\ref{t2c_amp} shows period--amplitude diagrams for T2Cs in GCs at optical and NIR wavelengths. For a relative comparison, we have also shown amplitudes of T2Cs in the LMC and the Galactic bulge. The optical amplitudes of BL Her show a marginal increase with period for up to 3 days, albeit with large scatter. The amplitudes of BL Her display an evidence of a decrease between 3 and 7 days. This suggests that these above HB variables display a wide range of light curve morphologies, even within a narrow period range. The amplitudes of W Vir increase sharply as a function of period between 5 and 20 days, while the amplitudes of RV Tau ($P>20$~days) decrease with increasing period. In the NIR bands, a significantly smaller scatter is seen in amplitudes for BL Her variables with periods less than 3 days. In this period range, the amplitudes seem to increase almost linearly as a function of period in the $K_s$-band. Similar trends to the optical bands are seen for W Vir and RV Tau at longer wavelengths. \citet{braga2020} investigated amplitude ratios for T2Cs and found a small but significant dependence on pulsation period, unlike RRL stars, which seem to have constant amplitude ratios for RRc and RRab stars \citep{braga2018, bhardwaj2020a}.

\subsection{Period--Luminosity Relations}

From the earlier studies of T2Cs in GCs based on photographic plates, it was known that these variables follow a PLR in optical bands \citep{harris1985}. Optical PLRs were also derived separately for short- and long-period T2Cs with a break at 10 days \citep{mcnamara1995}, and  two PLRs were even fitted to T2Cs in GCs, assuming their pulsations both in fundamental and overtone modes \citep{nemec1994}. However, modern data from time-domain surveys, in particular those in the Magellanic Clouds, revealed that T2Cs follow solid, optical band PLRs \citep{alcock1998, majaess2009, soszynski2008a}. Using high-precision {\it HST} photometric data for the two metal-rich clusters NGC 6388 and NGC 6441, which host most T2Cs, \citet{pritzl2003} derived well-defined and tight PLRs for T2Cs in $BVI$-bands. 

\textls[-5]{There is empirical evidence that T2Cs follow tight PLRs in GCs at NIR wavelengths~\citep{matsunaga2006}} similar to RRL stars. \citet{matsunaga2006} used NIR light curves of 46 T2Cs in 21 GCs to derive a linear PLR for BL Her, W Vir, and RV Tau variables. They used distances to individual GCs based on HB--metallicity relations, which also showed a consistency between the RRL and T2C distance scale. However, RV Tau variables in the Magellanic Clouds appear to be brighter than the PLR followed by the BL Her and W Vir stars \citep{ripepi2015}. \citet{bhardwaj2017a} also found an evidence of a linear relation for all T2Cs in the LMC. NIR observations of BL Her and W Vir in the bulge have also been used to investigate their PLRs and determine a distance to the galactic center \citep{bhardwaj2017b, braga2018a}.

\begin{figure}

\includegraphics[width=12 cm]{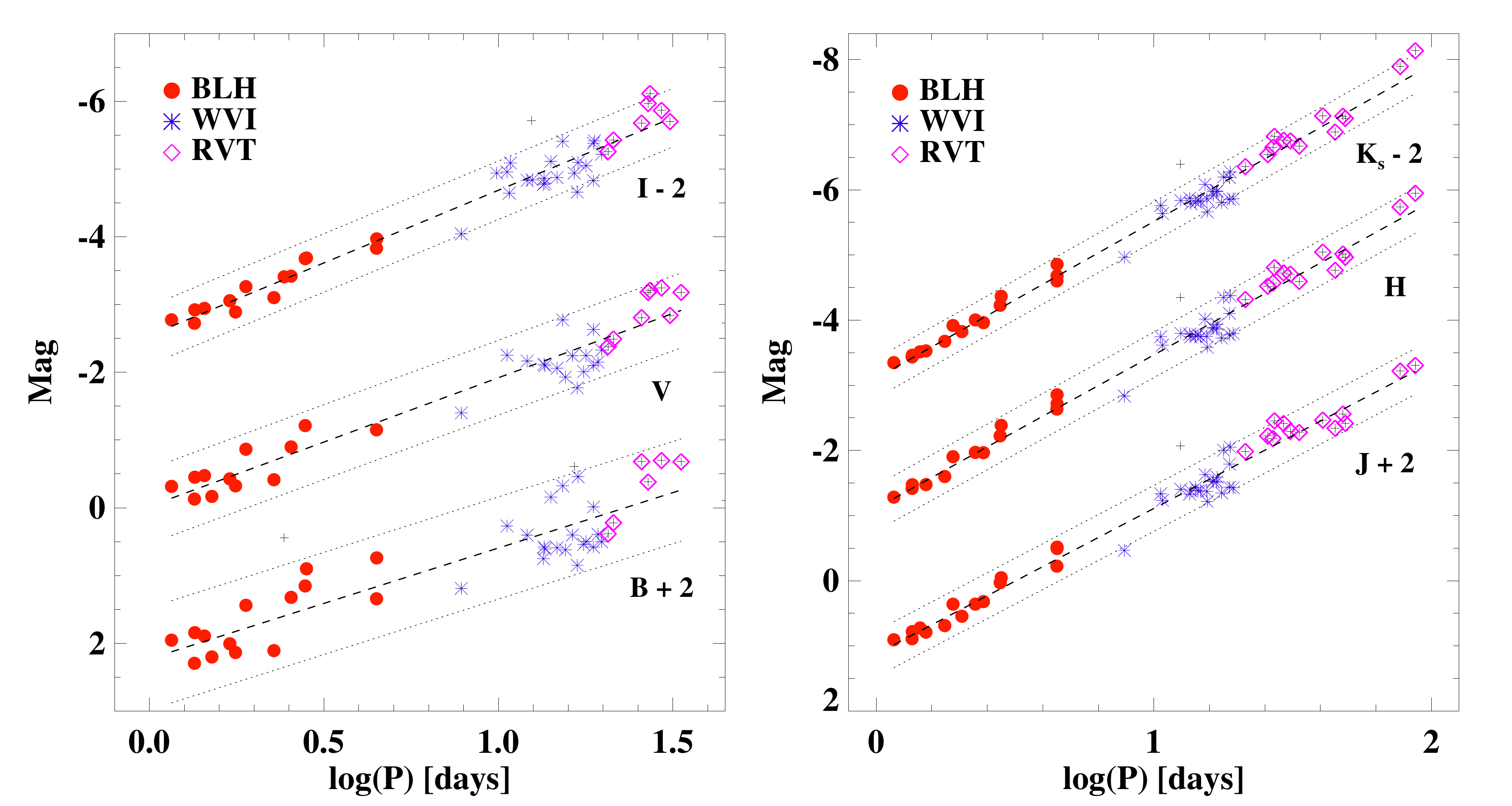}
\caption{{Optical} and NIR PLRs of T2Cs in GCs. Dashed lines represent best-fitting linear regression. Dotted lines display $\pm 2\sigma$ scatter in the best-fitting relation. Grey plus symbols represent outlier~variables. \label{t2c_plr}}
\end{figure}

Figure~\ref{t2c_plr} displays optical and NIR PLRs for T2Cs in GCs. A single slope is fitted to all three subclasses of T2Cs. The best-fitted PLRs are provided below: 
\begin{align}
\label{eq:rrl_plr_model}
\textrm{M}_B &= ~~~0.23~(0.13) -1.64~(0.13)\log(P) ~~~(\sigma=0.38,N=39)\nonumber\\
\textrm{M}_V &= -0.02~(0.10) -1.90~(0.09)\log(P) ~~~(\sigma=0.28,N=36)\nonumber\\
\textrm{M}_I &= -0.54~(0.07) -2.14~(0.07)\log(P) ~~~(\sigma=0.22,N=42)\nonumber\\
\textrm{M}_{J} &= -0.88~(0.06) -2.23~(0.05)\log(P) ~~~(\sigma=0.18,N=50)\nonumber\\
\textrm{M}_{H} &= -1.11~(0.06) -2.35~(0.05)\log(P) ~~~(\sigma=0.17,N=48)\nonumber\\
\textrm{M}_{K_S}&= -1.11~(0.05) -2.41~(0.04)\log(P) ~~~(\sigma=0.15,N=50)\nonumber
\end{align}

It is evident that the PLRs exhibit consistently smaller scatter at longer wavelengths.
The slope and zero-point of optical $BVI$ PLRs in GCs are statistically consistent with those derived in \citet{nemec1994} and \citet{pritzl2003}. The NIR PLRs are similar to those in \citet{matsunaga2006} because the T2C sample is the same with the addition of four more T2Cs (three in $\omega$ Cen and one in M15). In Figure~\ref{t2c_plr}, RV Tau seem to fall systematically on the brighter end of the best-fitting relation at optical wavelengths. However, there is no such evidence in the NIR data. 

There is theoretical and empirical evidence that the PLRs are minimally affected by the metal-abundances, in contrast 
to RRL stars. Theoretical predictions of \citet{di2007} and \citet{marconi2007} suggested that the metallicity effects are of the order of $0.05$ mag/dex on T2C PLRs. \citet{das2021} also computed BL Her models for a wide range of metal-abundances and derived multiband PLRs, which were found to be independent of metallicity. \citet{matsunaga2006} also found a negligible metallicity dependence on T2C PLRs at NIR wavelengths. We did not find any correlation of the residuals of PLRs with metallicity for T2Cs, implying a negligible dependence on metallicity. \citet{braga2020} used T2C PLRs in $\omega$ Cen to obtain a distance to the cluster that is consistent with independent methods, including those based on RRL stars. Since T2Cs and RRL follow the same PLRs in NIR bands, their combined PLRs can be used as an alternative to classical Cepheids for providing the calibration of the first rung of the extragalactic distance scale. 

\section{Summary}
\label{sec:sec7}

The article provides a review of recent advancements in the studies of the optical and NIR pulsation properties of RRL and T2Cs in GCs. While the modern time-domain surveys provide a vast amount of data at optical wavelengths, the $JHK_s$ time-series is still limited to only a few GCs, covering a statistically significant sample of RRL. Moreover, the crowded nature of GCs makes it difficult to obtain homogeneous photometry, even from time-domain surveys. We summarized light curve variations, color--magnitude and period--amplitude diagrams, and provided an empirical calibration of RRL and T2C PLRs at multiple wavelengths. The evolutionary scenario for RRL is well-understood, but the evolution of T2Cs, in particular, W Vir and RV Tau, is still debated in the literature. The pulsation properties of both RRL and T2Cs have now been explored in great detail at both the optical and NIR bands, which further strengthens their case for being potential primary distance indicators for the extragalactic distance scale. However, there still remain open questions related to the Oosterhoff dichotomy and the second parameter for the HB morphology in GCs. While some progress has been made in quantifying the metallicity effects on RRL and T2C PLRs, their absolute calibration is also being addressed with parallaxes from the {\it Gaia} mission. Our understanding of the impact of composition, metallicity, and helium effects in particular, on RRL and T2C pulsation properties will benefit from ongoing and upcoming large spectroscopic surveys. These cluster variables have played significant roles in our understanding of the evolution of stars and our Galaxy, and will continue enabling new insights into the mysteries of the Universe. 

\vspace{6pt}
\funding{{This research received no external funding.}}

\institutionalreview{{Not applicable}}

\informedconsent{{Not applicable}}


\dataavailability{{Not applicable}}


\conflictsofinterest{{The authors declare no conflict of interest.}}



\begin{adjustwidth}{-\extralength}{0cm}

\printendnotes[custom]
\reftitle{References}
%

\end{adjustwidth}
\end{document}